\documentclass[11pt, a4paper]{article}
\pdfoutput=1

\usepackage{amsmath,amssymb}
\usepackage{epsfig}
\usepackage{hyperref}
\usepackage{color}
\usepackage{cite}
\usepackage{multirow}

\usepackage{amsmath}
\usepackage{amsfonts}
\usepackage{amssymb}
\usepackage{graphicx}
\usepackage{epstopdf}
\usepackage{epsfig}
\usepackage{latexsym}
\usepackage{graphicx}
\usepackage{color}
\usepackage{amsmath,bm,amssymb}
\usepackage{cite}
\usepackage{slashed}
\usepackage{hyperref}
\usepackage{arydshln}

\definecolor{myred}{rgb}{0.7, 0, 0}
\definecolor{myblue}{rgb}{0, 0, 0.7}
\definecolor{mygreen}{rgb}{0.04, 0.7, 0.5}

\hypersetup{colorlinks,citecolor=myred,linkcolor=myblue,urlcolor=myblue,linktocpage=true}

\newcommand{\hhref}[1]{\href{http://arxiv.org/abs/#1}{arXiv:#1}}

\newcommand{\be}{\begin{equation}}
\newcommand{\ee}{\end{equation}}
\newcommand{\bea}{\begin{eqnarray}}
\newcommand{\eea}{\end{eqnarray}}
\newcommand{\eq}[1]{Eq.~(\ref{#1})}

\def\SO{\textrm{SO}}
\def\SU{\textrm{SU}}
\def\U{\textrm{U}}

\numberwithin{equation}{section}

\title{
\vspace{-2cm}
\vspace{3cm}
\bf \LARGE
EFT approach
 to the electron Electric Dipole Moment\\
  at the two-loop level
\vspace{.2cm}}
\date{}
\author{
{\large Giuliano Panico$^{a}$, Alex Pomarol$^{b,c}$, Marc Riembau$^{d}$}\\
[10mm]
\normalsize\itshape $^a$ Deutsches Elektronen-Synchrotron (DESY), 22607 Hamburg, Germany\\
\normalsize\itshape $^b$ IFAE and BIST, Universitat Aut\`onoma de Barcelona, 08193~Bellaterra,~Barcelona\\
\normalsize\itshape $^c$ Dept.~de~F\'isica, Universitat Aut{\`o}noma de Barcelona, 08193~Bellaterra,~Barcelona\\
\normalsize\itshape $^d$\, D\'epartment de Physique Th\'eorique, Universit\' e de Gen\`eve, Gen\`eve, Switzerland
}

\begin{document}

\begin{flushright}
DESY 18-185
\end{flushright}

{\let\newpage\relax\maketitle}
\begin{abstract}
\medskip
\noindent
The  ACME collaboration has  recently reported a new  bound on the electric dipole moment (EDM) of the electron, 
$|d_e|< 1.1 \times 10^{-29}\, {\rm e\cdot cm}$ at 90$\%$ confidence level,
reaching an unprecedented accuracy level.
This   can translate into  new relevant constraints 
 on  theories beyond the SM laying at  the TeV scale, 
 even when they contribute to the electron EDM
 at  the two-loop level.
We use the EFT approach to classify these corrections, 
presenting the  contributions to the anomalous dimension of the CP-violating dipole operators of the electron
up to the  two-loop level. 
Selection rules based on helicity and CP play an important role to simplify this analysis.
We use this result to  provide new  bounds on BSM with leptoquarks, extra Higgs, or  constraints in  sectors of the MSSM and composite Higgs models.
The new ACME bound pushes natural theories significantly more into fine-tune territory,  
unless  they have  a way to accidentally  preserve CP.
\end{abstract}

\newpage

\tableofcontents


\section{Introduction}

Electric dipole moments (EDM) provide  one of the best indirect probes for new-physics.
Since a non-zero EDM requires a violation of the CP symmetry, 
and the Standard Model (SM) contributions are accidentally highly suppressed, 
the EDM is an exceptionally  clean observable to uncover   beyond the SM (BSM)  physics.
Indeed,
if  BSM physics lies at the TeV scale, we expect new interactions 
and therefore new sources of  CP violation to be present,\footnote{As in the SM, we can expect that any parameter of the BSM that can be complex will be complex, providing unavoidably large new sources of CP violation.}
  inducing   sizable EDM to be observed in the near future.
  For this reason,  experimental bounds on the electron and neutron EDM  have provided the most substantial
  constraints on the best motivated BSM scenarios, such as supersymmetry or composite Higgs models.

The  ACME experiment has recently released   a new bound  on the electron EDM that improve by a factor $\sim 8.6$
their previous bound \cite{ACME}:
\be
|d_e|<1.1\cdot 10^{-29}\, {\rm e\cdot cm}\,.
\label{expbound}
\ee
This  unprecedented level of accuracy allows for a sensitivity to BSM effects even if they appear at the two-loop level.
Indeed,  using the rough estimate,
\be
\frac{d_e}{e}\simeq \left(\frac{g^2}{16\pi^2}\right)^2 \frac{m_e}{\Lambda^2}\,, 
\label{estimate}
\ee
we get from \eq{expbound} a bound on the scale of new-physics
 $\Lambda\gtrsim 2.5$ TeV,   being competitive with direct LHC searches.
It is therefore of crucial interest to understand how and which  BSM  sectors
 affect   the electron EDM up to the two-loop level, and which constraints  can be derived
 from the bound \eq{expbound}.

The purpose of the paper is to use  the Effective Field Theory (EFT) approach to
provide a classification of the leading BSM effects on the  EDM of the electron up to the two-loop level.
In the EFT approach BSM indirect effects are encoded in the Wilson coefficients of higher-dimensional SM operators.
At the loop level these Wilson coefficients can enter, via operator mixing,  into the renormalization of the 
CP-violating dipole operators responsible for the EDM. 
By  calculating  the anomalous dimensions of  these operators,
we can  provide all  log-enhanced  contributions to the EDM  coming from  new physics.

At the leading order in a $m^2_W/\Lambda^2$ expansion, 
the electron EDM arises from two dimension-6 operators, ${\cal O}_{eB}$ and ${\cal O}_{eW}$ (see below).
We will present   here the relevant anomalous dimensions of the imaginary part of the corresponding Wilson coefficients, $C_{eB}$ and  $C_{eW}$,   up to the two-loop level. 
In particular, we will provide the leading correction (either at the one-loop level or two-loop level)
of  the different  Wilson coefficients  $C_i$
to  the imaginary part of $C_{eB}$ and  $C_{eW}$.
We will see that due to selection rules, only few Wilson coefficients enter into the renormalization 
of $C_{eB}$ and  $C_{eW}$ at the one-loop or two-loop level.
Calculating these leading corrections will allow 
to extract  bounds on these Wilson coefficients  from the recent EDM measurement.

In addition,
we will also provide the most relevant  one-loop anomalous dimensions of the    dimension-8 operators
affecting the electron EDM. Although sub-leading in the $m^2_W/\Lambda^2$ expansion,
dimension-8 operators give contributions of order
\be
\frac{d_e}{e}\simeq \frac{g^2}{16\pi^2} \frac{m_e m_W^2}{\Lambda^4}\,, 
\ee
that  can also be  relevant as \eq{expbound} leads to the bound  $\Lambda\gtrsim 2$ TeV,  similar to those from \eq{estimate}.

 Our   results can be  useful   to derive  from  \eq{expbound}  new bounds     on  BSM particle  masses.
As an example, we will provide   bounds on BSM with leptoquarks or  extra Higgs, showing that
 we can exclude masses below  hundreds  of TeV.
We will also  present   constraints on new regions of the  parameter space of the MSSM,
as well as bounds on top-partners in composite Higgs models.
These bounds  can be better than those from present  and future direct searches at  the LHC, unless the BSM preserve CP.

\section{EDM of the electron in the EFT approach}

We are interested in calculating  new physics contributions to the electron EDM following the EFT approach.
This is a valid approximation whenever the new-physics scale $\Lambda$ is larger than the electroweak (EW) scale, such that 
 new-physics effects on the SM can be characterized by
the  Wilson coefficients $C_i(\mu)$ of  higher-dimensional SM operators.
Assuming lepton number conservation,
the leading effects arise from  dimension-6 operators
\be
\Delta {\cal L}=\sum_i\frac{C_i(\mu)}{\Lambda^2}{\cal O}_i\,,
\ee
where the Wilson coefficients are  induced at   the new-physics scale,  $C_i(\mu=\Lambda)$,
and  must be evolved  via the renormalization group equations (RGEs) down to the relevant physical scale at which  the measurement takes place.
Since the   Wilson coefficients  mix via loop effects in the RGEs, 
precise measurements, such as the EDM,  can be sensitive to different Wilson coefficients
induced by different sectors of the BSM.

The EDM of the electron is measured at low-energies $\mu\ll m_e$,
and can then be extracted in the EFT approach
from  the coefficient of the operator
\be
-\frac{i}{2}d_e(\mu)\, \bar e\sigma_{\mu\nu}\gamma_5 eF^{\mu\nu}\,,
\label{edmdef}
\ee
evaluated at  the electron mass,
\be
d_e=d_e(\mu=m_e)\, ,
\ee
 where $F_{\mu\nu}$ is the field-strength of the photon.
The RG evolution   of $d_e(\mu)$  from $\Lambda$ to $m_e$
must be computed in the EFT made with the states lighter  than $\mu$.
 This means that from the new-physics  scale $\Lambda$ down to the EW scale
we must use the SM EFT, while below
 the EW scale we must use the effective theory including only light SM fermions, gluons and photons.
Let us  start  discussing the contributions to $d_e(\mu)$
in the SM EFT.

\subsection{SM EFT basis}

We will work mainly within the Warsaw basis \cite{Grzadkowski:2010es}, as 
the loop operator mixing  is simpler in this basis due to the presence of 
many non-renormalization  results.
Nevertheless, we will  make  two changes
in the four-fermion  operators of the Warsaw  basis.
In particular,  we will make the replacement\footnote{These operators are related to the ones in the Warsaw basis \cite{Grzadkowski:2010es} by  Fierz identities, namely ${\cal O}^{(3)}_{lequ} = - 8 {\cal O}_{luqe} - 4 {\cal O}^{(1)}_{lequ}$ and
${\cal O}_{le}=-2 {\cal O}_{le\bar e'\bar l'}$.}
\bea
{\cal O}^{(3)}_{lequ}=(\bar L_L^a\sigma_{\mu\nu}e_R) \varepsilon_{ab} (\bar Q_L^b \sigma^{\mu\nu}u_R) &\ \to\ &
{\cal O}_{luqe}=(\bar L_L^a u_R) \varepsilon_{ab}  (\bar Q_L^b e_R)\,,\\
{\cal O}_{le}=(\bar L_L\gamma^\mu L'_L)(\bar e'_R\gamma_\mu e_R)
  &\ \to\ &
   {\cal O}_{le\bar e'\bar l'}=(\bar L_L e_R)(\bar e'_RL_L')\,,
\eea
where $a,b$ denote the $\SU(2)_L$ doublet indices,
and  $L_L$ and $e_R$ denote only the first generation lepton multiplets, while $L'_L$ and $e'_R$  the second and third generation ones.\footnote{Notice that in our formulae we will suppress the fermion generation indices, except in the cases in which they cannot be straightforwardly reconstructed from the context.}
We also relabel the operator
\be
{\cal O}_{ledq}=(\bar L_L^a e_R)(\bar d_R Q_{L\,a})\equiv{\cal O}_{le\bar d\bar q}\,.
\ee
Our labeling is to make clear that there are two types of operators
${\cal O}_{\psi\psi\psi\psi}$ and ${\cal O}_{\psi\psi\bar \psi \bar \psi}$ that in
Weyl notation   are respectively  $\psi^4$ of total helicity 2 and 
$\psi^2\bar\psi^2$ of total helicity zero. 
 As we will see in the following, the helicity of the operator  plays a crucial role in understanding 
the properties of the operator mixing at the loop level \cite{Cheung:2015aba}.
In Dirac notation these two type of operators could also  be written   (after Fierzing)  respectively 
as operators  of type $(\bar \Psi\gamma_\mu\Psi)(\bar \Psi\gamma^\mu\Psi)$
 and  of type $(\bar \Psi_L \Psi_R)(\bar \Psi_L \Psi_R)$.
In this case, for example, we would have ${\cal O}_{le\bar d\bar q}=-(\bar L_L^a\gamma_\mu Q_{L\,a})(\bar  d_R\gamma^\mu e_R)/2$.

\renewcommand{\arraystretch}{1.4}
\begin{table}[t]
\begin{center}
\begin{tabular}{|l|}
\multicolumn{1}{c}{tree level}\\
\hline
${\cal O}_{eW}=(\bar L_L\sigma^a \sigma^{\mu\nu} e_R)\, H W^a_{\mu\nu}$\\
${\cal O}_{eB}=(\bar L_L \sigma^{\mu\nu} e_R)\, H B_{\mu\nu}$\\
[.5ex]   \hline
\multicolumn{1}{c}{\rule{0pt}{2.em}1-loop}\\
\hline  
  ${\cal O}_{luqe}=(\bar L_L u_R)(\bar Q_L e_R)$\\
  ${\cal O}_{W\widetilde W}=|H|^2\, W^{a\, \mu\nu}\widetilde W^a_{\mu\nu}$\\
${\cal O}_{B\widetilde B}=|H|^2\, B^{\mu\nu}\widetilde B_{\mu\nu}$\\
${\cal O}_{W\widetilde B}=(H^\dagger\sigma^aH)\, W^{a\, \mu\nu}\widetilde B_{\mu\nu}$\\
\hdashline
${\cal O}_{\widetilde W} = \varepsilon_{abc} \widetilde W^{a \nu}_\mu W^{b\rho}_\nu W^{c\mu}_\rho$\\
[.5ex]
\hline
 \end{tabular}\hspace{10mm}
 \begin{tabular}{|l|}
 \multicolumn{1}{c}{2-loop}\\
 \hline
  ${\cal O}^{(1)}_{lequ}=(\bar L_L e_R)(\bar Q_L u_R)$\\
  ${\cal O}_{e'W} = (\bar L'_{L}\sigma^a \sigma^{\mu\nu} e'_{R})\, H W^a_{\mu\nu}$\\
  ${\cal O}_{e'B}=(\bar L'_{L} \sigma^{\mu\nu} e'_{R})\, H B_{\mu\nu}$\\
  ${\cal O}_{uW}=(\bar Q_L\sigma^a \sigma^{\mu\nu} u_R)\, \widetilde H W^a_{\mu\nu}$\\
  ${\cal O}_{uB}=(\bar Q_L \sigma^{\mu\nu} u_R)\, \widetilde H B_{\mu\nu}$\\
  ${\cal O}_{dW}=(\bar Q_L\sigma^a \sigma^{\mu\nu} d_R)\, H W^a_{\mu\nu}$\\
  ${\cal O}_{dB}=(\bar Q_L \sigma^{\mu\nu} d_R)\, H B_{\mu\nu}$\\
  \hdashline
  ${\cal O}_{le \bar d\bar q}= (\bar L_L e_R)(\bar d_R Q_L)
$\\
  ${\cal O}_{le\bar e'\bar l'}=(\bar L_L e_R)(\bar e'_R L'_L)
$\\
  ${\cal O}_{y_e}=|H|^2 \bar L_L e_RH$\\
[.5ex]\hline
 \end{tabular}
\caption{\it Operators involved in our analysis. Top-left: Operators contributing to the electron EDM at tree-level. Bottom-left: Operators  contributing to the electron EDM  at the one-loop level via mixing. Right: Operators contributing to the electron EDM at the two-loop level via mixing. In this table we denoted by $L_L$ and $e_R$ only the first generation lepton multiplets, while $L'_L$ and $e'_R$ denote the second and third generation ones.}
\label{operatortable}
\end{center}
\end{table}

\subsection{Tree-level contributions}

At tree-level there are only two dimension-6 operators that contribute to the electron EDM, namely the dipole operators ${\cal O}_{eW}$ and ${\cal O}_{eB}$, given in Table~\ref{operatortable}.
We have
\be
d_e(\mu)=\frac{\sqrt{2}v}{\Lambda^2}{\rm Im}\left[s_{\theta_W}\, C_{eW}(\mu)-c_{\theta_W}\, C_{eB}(\mu)\right]\,,
\ee
where we defined $v\simeq 246$ GeV as the Higgs VEV, and
$s_{\theta_W}\equiv \sin\theta_W$ with $\theta_W$ the weak angle (similarly for the other trigonometric functions).
Notice that contributions to the electron EDM arised only from the imaginary part of $C_{eW,eB}$.
For this reason, we will only be interested in loop contributions to the dipole operators that can generate 
nonzero ${\rm Im}[C_{eW,eB}$].

\subsection{One-loop effects}\label{sec:one-loop}

At the one-loop level, however, other dimension-6 operators can mix with the dipole operators ${\cal O}_{eW}$ and ${\cal O}_{eB}$, giving
 a contribution to $d_e$.
Selection rules, mainly based on helicity arguments, 
dictate that only few operators
contribute at the one-loop order to the anomalous dimension of 
the Wilson coefficients  $C_{eW}$ and/or $C_{eB}$, as has been argued in Refs.~\cite{Elias-Miro:2014eia,Cheung:2015aba} based on the analysis
of Ref.~\cite{Alonso:2014rga}.
The relevant selection rules for our analysis are given in Table~\ref{selectionrule},\footnote{There is an exemption to these selection rules when the  pair of Yukawas $y_uy_e$ or $y_uy_d$
  is involved in the loop, as this  can induce  a mixing from $\psi^2\bar \psi^2$  to $\psi^4$   \cite{Elias-Miro:2014eia,Cheung:2015aba}.   This can only affect the EDM operators at the two-loop level and will be discussed later.}
 where
we use Weyl notation, and denote  with $F$  any SM field strength, with $\psi$ any Weyl fermion
and  with $H$ any Higgs insertion.
The ${\cal O}_{eW}$ and ${\cal O}_{eB}$ operators are of type $H\psi^2F$
and can then only  receive contributions from operators of type $\psi^4$, $F^3$ and $H^2F^2$ of total helicity  $\geq 2$.

There are four dimension-6 operators of type $\psi^4$, but
only two  contain two leptons, 
${\cal O}_{luqe}$ and ${\cal O}^{(1)}_{lequ}$ given in Table~\ref{operatortable}.
The second one, however, after closing the quark loop, can only give rise to the Lorentz-singlet structure $\bar L_L e_R$,
and therefore cannot contribute to the electron dipoles.
Hence only ${\cal O}_{luqe}$ contributes at the one-loop level to the anomalous dimension of 
${\cal O}_{eW}$ and ${\cal O}_{eB}$.\footnote{\label{footnote} As explained in Ref.~\cite{Elias-Miro:2014eia}, this is easily seen in Weyl notation where the dipole operator is  $\propto  L_\alpha E_\beta F^{\alpha\beta}$ with $L_\alpha$  and $E_\beta$ being respectively  the ${\rm SU(2)}_L$  doublet and singlet Weyl electron.  Therefore, only four-fermion operators containing  $L_\alpha E_\beta$ 
(antisymmetric under $\alpha\leftrightarrow \beta$) can contribute at the loop level  to the dipole. The  only one is $L_\alpha E_\beta U^\alpha Q^\beta$ that corresponds to ${\cal O}_{luqe}$ in Dirac notation.
Notice that this argument applies also to one-loop finite parts.}
This contribution is given by
\be
\frac{d}{d\ln \mu}
\left(\begin{array}{c} C_{eB}\\ C_{eW}\end{array}\right)
=\frac{y_ug}{16\pi^2}
\left(
\begin{array}{c} -\frac{1}{2}  t_{\theta_W} N_c (Y_Q+Y_u) \\
\frac{1}{4} N_c
\end{array}\right)
C_{luqe}\,,
\label{eq:cyuyetoWdipolerge}
\ee
where $Y_f$ refers to the hypercharge of the fermion $f$ ($Y_Q=1/6$, $Y_u=2/3$ and $Y_d=-1/3$), and $N_c=3$
is the number of QCD colors.
Since we can work in a basis where the Yukawa matrix $y_u$ is diagonal, the renormalization of the imaginary part of 
$C_{eW,eB}$ from \eq{eq:cyuyetoWdipolerge}  only arises from the imaginary part of $C_{luqe}$.

\renewcommand{\arraystretch}{1.4}
\begin{table}[t]
\begin{center}
\hspace{-13mm}$\curvearrowright$
\hspace{25mm}
$\curvearrowright$\\
\begin{tabular}{|l|}\hline
$F^3$\\
[.5ex]
\hline
\end{tabular}
\ $\rightarrow$\ \ \begin{tabular}{|l|}\hline
  $H\psi^2F\,,\ \psi^4\,,\ H^2F^2$\\
[.5ex]\hline
 \end{tabular}
\caption{\it Selection rules \cite{Elias-Miro:2014eia,Cheung:2015aba} for the mixing  at the one-loop level between the different types of dimension-6 operators  (in Weyl notation).}
\label{selectionrule}
\end{center}
\end{table}

A second type
of  operators, involving SM bosons, are $H^2 F_{\mu\nu}\widetilde F^{\mu\nu}$.
There are three operators of this type 
in the SM, presented  in the bottom-left of Table~\ref{operatortable}.
All of them contribute to the EDM at the one-loop level \cite{Alonso:2013hga}:
\be
\frac{d}{d\ln \mu}{\rm Im}
\left(\begin{array}{c} C_{eB}\\ C_{eW}\end{array}\right)
=-\frac{y_eg}{16\pi^2}
\left(
\begin{array}{ccc} 
0&\ \ 2t_{\theta_W}(Y_L+Y_e)&\ \ \frac{3}{2}\\  1&0&t_{\theta_W}(Y_L+Y_e)
\end{array}\right)
\left(\begin{array}{c} C_{W\widetilde W}\\ C_{B\widetilde B}\\ C_{W\widetilde B}\end{array}
\right)\,.
\label{r2}
\ee
It is instructive to  write \eq{r2} in a more physically oriented way, by relating the contributions to the EDM to
those to the CP-violating  Higgs couplings $h\gamma\gamma$, $h\gamma Z$, and to the anomalous triple gauge coupling $\delta\tilde\kappa_\gamma$, defined as
\be
 \frac{v h}{\Lambda^2}
 \left( \tilde{\kappa}_{\gamma\gamma}F_{\mu\nu}\widetilde{F}^{\mu\nu} +2 \tilde{\kappa}_{\gamma Z} 
 {F}_{\mu\nu} \widetilde Z^{\mu\nu}\right)
 + ie \delta\tilde{\kappa}_\gamma W^+_\mu W^-_\nu\widetilde{F}^{\mu\nu}\,.
\ee
We have
\bea
\tilde{\kappa}_{\gamma\gamma} &=& c^2_{\theta_W}C_{B\tilde{B}}+s^2_{\theta_W}C_{W\widetilde{W}}-c_{\theta_W}s_{\theta_W}C_{W\widetilde{B}}\,,\nonumber\\
\tilde{\kappa}_{\gamma  Z} &=& c_{\theta_W}s_{\theta_W}\left( C_{W\widetilde{W}}-C_{B\widetilde{B}}\right)
-\frac 12(c_{\theta_W}^2-s_{\theta_W}^2)C_{W\widetilde{B}}\,,\nonumber\\
\delta\tilde{\kappa}_\gamma &=& \frac{1}{t_{\theta_W}}\frac{v^2}{\Lambda^2} C_{W \widetilde{B}}\,,
\eea
that, using \eq{r2}, leads  to
\be
\frac{d}{d\ln \mu} d_e(\mu)=
\frac{ e}{8\pi^2} \frac{m_e}{\Lambda^2}
\left[ 4 Q_e\tilde{\kappa}_{\gamma\gamma}
 -\frac{4}{s_{2\theta_W}}\left(\frac{1}{2}+2Q_e s^2_{\theta_W}\right)
  \tilde{\kappa}_{\gamma Z} + \frac{\Lambda^2}{v^2}\delta\tilde{\kappa}_\gamma\right]\,,
\label{total}
\ee
where $Q_e\equiv -1/2+Y_L=Y_e = -1$ is the electric charge of the electron.
Due to the approximate accidental  cancellation in the electron vector coupling to the $Z$, $ (1/2+2Q_es^2_{\theta_W})\sim 0.04$, the main contribution to the EDM comes
from $\tilde{\kappa}_{\gamma\gamma}$ and $\delta\tilde{\kappa}_\gamma$.
In fact, this second contribution is often found  to be small in many BSM scenarios, such as the MSSM or composite Higgs models, as we will see later. In these models the contribution from $\tilde{\kappa}_{\gamma\gamma}$ is the dominant one.
This allows in many cases  to give a direct relation between the electron EDM and the CP-violating Higgs coupling to photons.

Another class of operators that can in principle mix at the one-loop level with the electron dipole operators
are other type of dipole operators $H\psi^2F$, for example, those involving other fermions in the SM.
It is easy to see, however,  that there are no possible Feynman diagrams from quark dipole operators
contributing to the electron EDM at the one-loop level.
For  dipole operators involving other SM leptons, for example, $He\mu F$,
these contributions are also absent at the one-loop level.
Indeed, since we can work in a basis where the SM lepton Yukawa matrix $y_e$ is diagonal,
none of these operators can affect the electron EDM at the one-loop level. 
Below we will see that there can be, however,
contributions at the two-loop level.

Finally, there are operators of type $F^3$ that could potentially  mix with the dipole operators.
In the SM there are two of these operators, either with gluons (${\cal O}_{\widetilde G}$) or $W^a$ bosons
(${\cal O}_{\widetilde W}$). The ${\cal O}_{\widetilde G}$ operator obviously can not give corrections to the electron dipole operators at one-loop or two-loop level, as there are no possible Feynman diagrams at these orders.
On the other hand, the ${\cal O}_{\widetilde W}$ operator can contribute to the ${\cal O}_{eW}$ dipole
at one loop. It turns out, however, that this contribution is finite, so it does not induce a running for the
dipole operator. The finite contribution can be readily computed in dimensional regularization, leading
to the result~\cite{Boudjema:1990dv}\footnote{Notice that the result depends on the regularization procedure used to compute
the one-loop integral~\cite{Boudjema:1990dv}. It has been argued in Ref.~\cite{Gripaios:2013lea} that, for this computation,
dimensional regularization provides the only sensible regularization procedure within the EFT framework.}
\begin{equation}
{\rm Im}[C_{eW}]= \frac{3}{64 \pi^2} y_e g^2 C_{\widetilde W}\,.
\label{eq:cwfinite}
\end{equation}

\subsection{Two-loop effects}

At the two-loop level, more dimension-6 operators can contribute to the electron EDM by mixing with the dipoles
${\cal O}_{eW}$ and ${\cal O}_{eB}$.
We remark that we are only interested
in calculating two-loop effects for those Wilson coefficients
that did not mix with the EDM operators at the loop level. We are interested in extracting 
bounds to their corresponding Wilson coefficients and therefore
we are only interested in calculating the leading correction to the EDM.
For Wilson coefficients affecting the EDM already at the one-loop level, such as $C_{luqe}$, 
the two-loop corrections would only  provide a small correction to their bound.

\begin{figure}
\centering
\includegraphics[width=.99\textwidth]{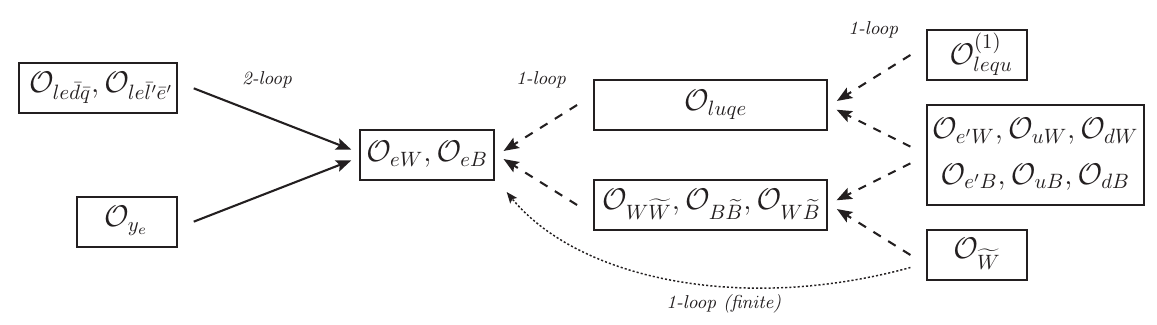}
\caption{\it  Corrections to the  electron EDM (imaginary part of $C_{eW,eB}$) induced up to the 2-loop level. The dashed and solid arrows denote
mixing at 1-loop and 2-loop order respectively. On the right we list the operators that generate
contributions enhanced by a double logarithm (showing the 1-loop mixing patterns that generate it),
whereas on the left we list  operators giving rise to a single logarithm.}
\label{fig:2-loop_running}
\end{figure}

New dimension-6 operators can contribute to the electron EDM by mixing with the dipoles
${\cal O}_{eW}$ and ${\cal O}_{eB}$  in two different ways. 
Either by mixing at the one-loop level with the operators we discussed in the previous section, ${\cal O}_{luqe}$ and  ${\cal O}_{V\tilde V}$ ($V=W,B$), that contribute at the one-loop level to the dipoles,
or by direct two-loop contribution to the  anomalous dimension of 
${\cal O}_{eW}$ and ${\cal O}_{eB}$ (see Table~\ref{fig:2-loop_running}).

The first case can potentially give larger corrections, as in the leading-log approximation,
 they will contain  two logarithms, i.e.~$\propto \ln^2 (\Lambda^2/m_W^2)$.
From the selection rules of Table~\ref{selectionrule}, we see that only two classes of operators can contribute at this order.
One is given by the $\psi^4$ operators that could not generate an electron dipole at the one-loop
due to the absence of Feynman diagrams, namely the ${\cal O}^{(1)}_{lequ}$ operator. The second class is given by dipole operators involving the second and third
lepton generations, ${\cal O}_{e'W}$ and ${\cal O}_{e'B}$, or the quarks, ${\cal O}_{uW}$, ${\cal O}_{uB}$,
${\cal O}_{dW}$ and ${\cal O}_{dB}$.

Notice that, as we pointed out before, there is an exception 
to the selection rules of Table~\ref{selectionrule}, corresponding to 
a possible mixing of  $\bar \psi^2 \psi^2$ operators into $\psi^4$ 
when the pair of Yukawas either $y_uy_e$ or $y_uy_d$
  is involved in the loop  \cite{Elias-Miro:2014eia,Cheung:2015aba}.   
Nevertheless,  
by working in the basis in which the lepton and up-type quark Yukawa matrices are real and diagonal,
one can easily find that there are not $\bar \psi^2 \psi^2$ operators  contributing to the imaginary part of  ${\cal O}_{luqe}$ at the one-loop level.
Indeed, in this basis $y_uy_e$ is real and diagonal,
and  the only $\bar \psi^2 \psi^2$ operators 
that could contribute to ${\cal O}_{luqe}$
are the ones involving two electron fields and two same-generation quarks. The Wilson coefficients of these operators are necessarily real and do not induce CP-violating effects.

Therefore  the one-loop mixing pattern and RGEs are  the following.
The ${\cal O}_{lequ}$ operator can mix with ${\cal O}_{luqe}$ at the one-loop level \cite{Alonso:2013hga}:
\be
\frac{d}{d\ln \mu}  C_{luqe}=\frac{g^2}{16\pi^2}\left[ 4(Y_L+Y_e)(Y_Q+Y_u)t^2_{\theta_W}-3 \right] C^{(1)}_{lequ}\,.
\label{ltol}
\ee
The dipole operators, on the other hand, mix with the ${\cal O}_{luqe}$ operator, \cite{Alonso:2013hga}\footnote{The heavy lepton dipole operators induce a running for ${\cal O}_{luqe}$ at one loop. However in the basis with diagonal lepton Yukawa's
they contribute only to the ${\cal O}_{luqe}$ involving heavy leptons, which then does not contribute to the
electron dipoles at one loop.}
\begin{equation}
\frac{d}{d\ln \mu}  C_{luqe} = \frac{g\, y_e}{16\pi^2}\Big[- 8 t_{\theta_W} (Y_L + Y_e) C_{uB}
+ 12 C_{uW}\Big]\,,
\end{equation}
as well as with ${\cal O}_{V\widetilde V}$ operators   \cite{Elias-Miro:2013gya}:
\begin{equation}
\frac{d}{d\ln \mu}  C_{W\widetilde W} = -\frac{2 g}{16 \pi^2}{\rm Im}\Big[y_{e'}\, C_{e'W}
+ y_{u} N_c\, C_{uW} + y_{d} N_c\, C_{dW}\Big]\,,
\end{equation}
\begin{equation}
\frac{d}{d\ln \mu}  C_{B\widetilde B} = -\frac{4 g'}{16 \pi^2}{\rm Im}\Big[y_{e'} (Y_L + Y_e)\, C_{e'B}
+ y_{u} N_c (Y_Q + Y_u)\, C_{uB} + y_{d} N_c (Y_Q + Y_d)\, C_{dB}\Big]\,,
\end{equation}
\begin{eqnarray}
\frac{d}{d\ln \mu}  C_{W\widetilde B} &=& -\frac{2 g}{16 \pi^2}{\rm Im}\Big[
2t_{\theta_W}\big(y_{e'} (Y_L + Y_e)\, C_{e'W}
- y_{u} N_c (Y_Q + Y_u)\, C_{uW} + y_{d} N_c (Y_Q + Y_d)\, C_{dW}\big)\nonumber\\
&& \hspace{4.75em}+\,y_{e'}\, C_{e'B}
- y_{u} N_c\, C_{uB} + y_{d} N_c\, C_{dB}\Big]\,.
\end{eqnarray}
Flavor indices are easily understood as we can always work with diagonal Yukawa matrix $y_e$ and either
$y_u$ or $y_d$.

The operator $\mathcal{O}_{\widetilde{W}}$ also enters at two loops via renormalization of the ${\cal O}_{V\tilde V}$ operators \cite{Alonso:2013hga},
\begin{equation}
\frac{d}{d\ln \mu}  C_{W\widetilde W} = - \frac{1}{16\pi^2} 15 g^3 C_{\widetilde{W}}\ ,\hspace{1cm}
\frac{d}{d\ln \mu}  C_{W\widetilde B} = + \frac{1}{16\pi^2} 6 {g^\prime} g^2 Y_H  C_{\widetilde{W}}\, .
\end{equation}
This leads to a two loop, double log contribution to the electron EDM, to be compared with the finite contribution at one loop in \eq{eq:cwfinite}. The two loop contribution becomes comparable to the one-loop one already for $\Lambda \sim 10~\text{TeV}$.

\begin{figure}[t]
\centering
\includegraphics[width=.3\textwidth]{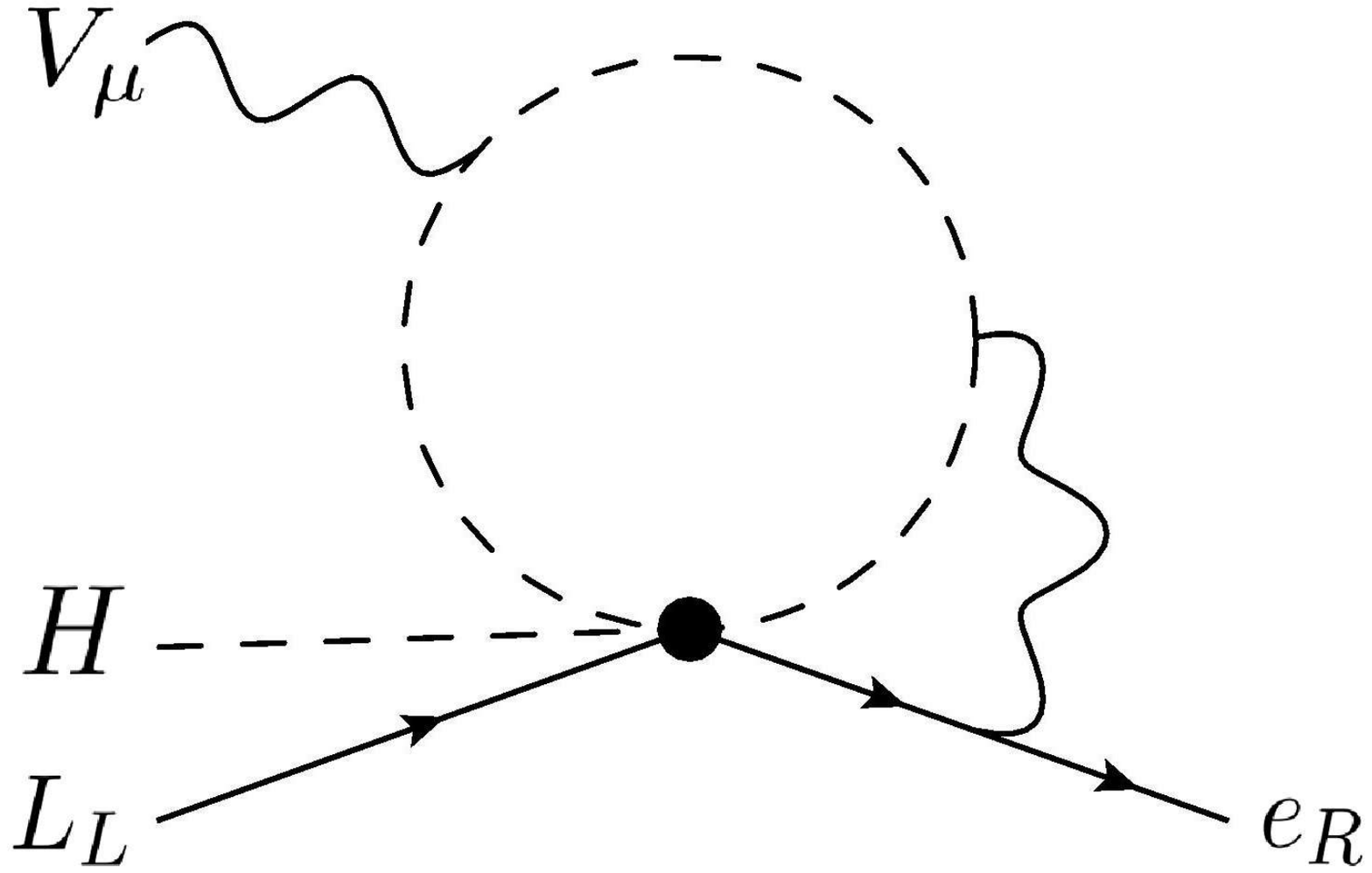}
\hskip1.5cm
\includegraphics[width=.3\textwidth]{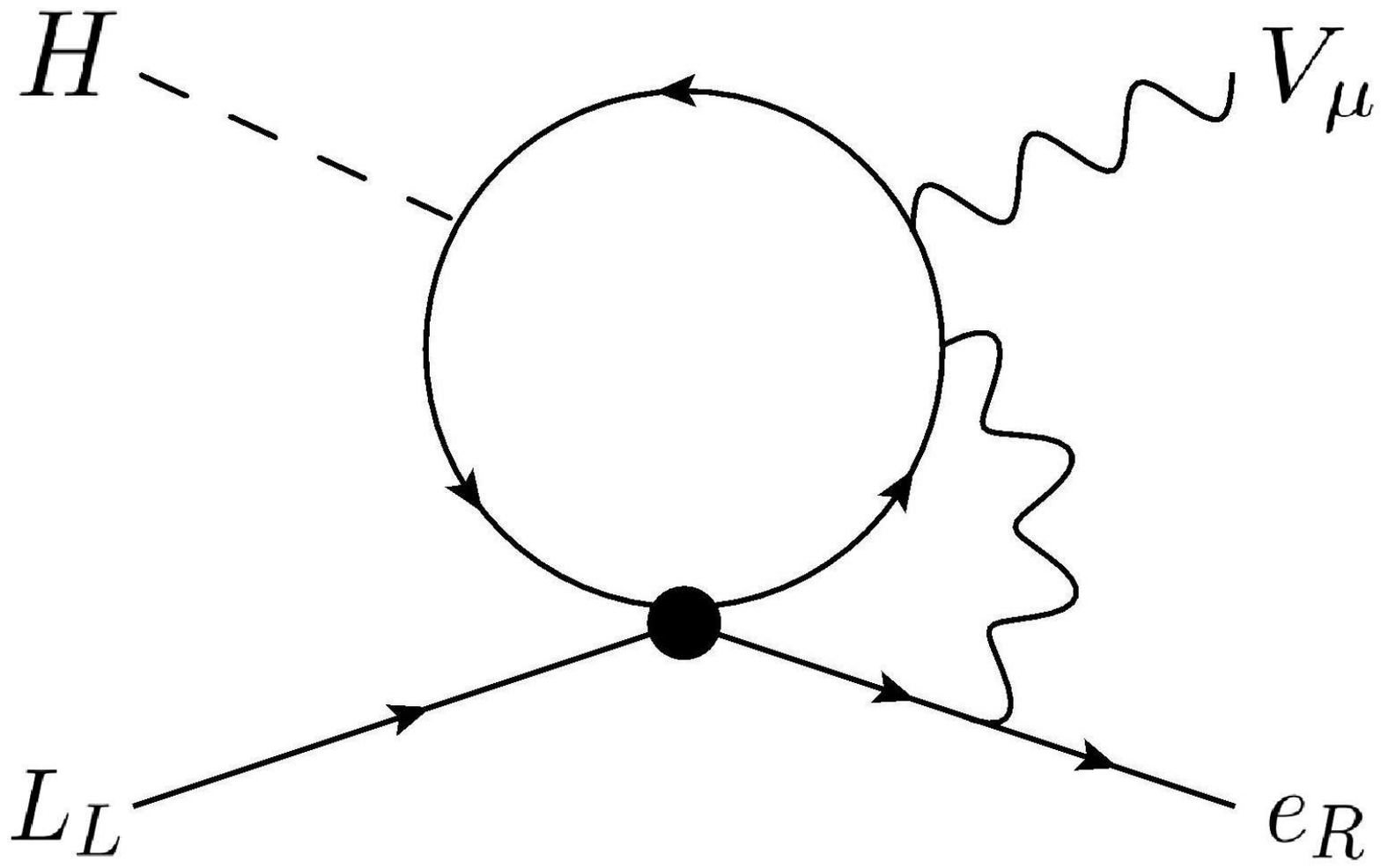}

\caption{\it Feynman contributions to the anomalous dimension of ${\cal O}_{eW}$ and ${\cal O}_{eB}$
at the two-loop level from
${\cal O}_{y_e}$ (left) and ${\cal O}_{le\bar d\bar q}$,${\cal O}_{le\bar e'\bar l'}$  (right).
}
\label{feynman2loop}
\end{figure}

Let us now discuss those dimension-6 operators that can directly  contribute at the two-loop level
to the anomalous dimension of the electron dipole operators. In fact,  we are only interested in the EDM, i.e. the imaginary part of the the dipole operators, and therefore  only complex Wilson coefficients can contribute, as the SM interactions preserve CP up to small Yukawa couplings that we neglect. This reduces the list of possible dimension-6 operators to those to the right of Table~\ref{operatortable}. 
For example,  operators of the type $\bar\Psi\gamma_\mu\Psi H^\dagger D^\mu H$ are Hermitian (and then have real Wilson coefficients) unless the two fermions involved are different, meaning that they must involve different flavors.
But since in the SM we can work in the basis where $y_e$ and either $y_u$ or $y_d$ are diagonal,
we cannot draw any two-loop Feynman diagram contributing to the electron EDM operators.

Similar conclusions can be obtained for four-fermion operators, except for those in Table~\ref{operatortable},
namely ${\cal O}_{le\bar d\bar q}$ and ${\cal O}_{le\bar e'\bar l'}$. There is however an important
subtlety related to these operators, which results in an ambiguity in the determination of their contributions
to the electron EDM.

Within the basis we are using, in which ${\cal O}_{le\bar d\bar q}$ and ${\cal O}_{le\bar e'\bar l'}$ are written as the product
of scalar currents, it is simple to check that the $1$-loop contributions to the electron EDM trivially vanish
due to the tensor structure. However, through a Fierz rearrangement, ${\cal O}_{le\bar d\bar q}$ and ${\cal O}_{le\bar e'\bar l'}$ can also be rewritten in the form $(\bar\Psi\gamma_\mu\Psi)(\bar\Psi\gamma^\mu\Psi)$, i.e.
as a product of vector currents. With this choice, if dimensional regularization (in particular the $\overline{\rm MS}$ scheme)
is used to compute the contributions to the electron EDM, a finite $1$-loop effect is found.
The origin of this contribution is related to the presence of additional four-fermion interactions involving multiple
gamma matrices that are generated at intermediate steps of the calculation. These are known as ``evanescent operators''
(for a review, see for example \cite{Herrlich:1994kh}).
The coefficients of these interactions carry an $\epsilon = 4-d$ factor, but they can give finite effects in the presence of $1/\epsilon$ poles. As an example, we report the contributions to the electron EDM induced
by the ${\cal O}_{\ell e} = C_{\ell e} (\bar L_L \gamma^\mu L'_L)(\bar e'_R \gamma_\mu e_R)$ operator (see for instance~\cite{Crivellin:2013hpa, Frigerio:2018uwx})
\begin{equation}
\frac{d_e}{e} = 2\frac{m_{\ell'}}{16 \pi^2} {\rm Im}\,C_{\ell e}\,.
\end{equation}
A similar result is obtained for the ${\cal O}_{le\bar d\bar q}$ operator in vector-current form.

Summarizing the above discussion, one finds that the contributions from the ${\cal O}_{le\bar d\bar q}$ and ${\cal O}_{le\bar e'\bar l'}$ $4$-fermion operators crucially depend on the choice of the operator basis and on the regularization procedure.
This, in turn, can affect the matching from a UV model. In particular, finite $1$-loop contributions can be shifted from the
matching to the $4$-fermion operators into the EDM operators ${\cal O}_{eB}$ and ${\cal O}_{eW}$ and vice versa.\footnote{Analogous ambiguities are present in the case of $4$-fermion contributions to the magnetic dipole moments.
For instance see the analysis of the $b \rightarrow s \gamma$ transitions in ref.~\cite{Ciuchini:1993ks}.}
As already mentioned, for our analysis we choose the scalar-current form for the $4$-fermion operators,
$(\bar L_L e_R)(\bar \Psi \Psi)$. Therefore no finite one-loop contributions arise for the electron EDM from the
${\cal O}_{le\bar d\bar q}$ and ${\cal O}_{le\bar e'\bar l'}$ operators. As we will see in sec.~\ref{sec:heavy_higgs}, this choice
is particularly convenient for studying UV models including heavy Higgs-like states, in which case only scalar-current
$4$-fermion operators are obtained from the matching.

We have then only ${\cal O}_{y_e}$,  ${\cal O}_{le\bar d\bar q}$ and ${\cal O}_{le\bar e'\bar l'}$
giving a  two-loop mixing with the electric dipole operators.
From ${\cal O}_{y_e}$   (see left-hand side of Fig.~\ref{feynman2loop}), we obtain
\be
\frac{d}{d\ln \mu}
\left(\begin{array}{c} C_{eB}\\ C_{eW}\end{array}\right)
=\frac{g^3}{(16\pi^2)^2}\frac{3}{4}
\left(
\begin{array}{c}
t_{\theta_W} Y_H + 4 t^3 _{\theta_W}Y_H^2 (Y_L +Y_e)
\\
\frac{1}{2} + \frac{2}{3} t^2 _{\theta_W} Y_H (Y_L+Y_e)
\end{array}
\right)
C_{y_e}\,,
\label{eq:r3}
\ee
while  from ${\cal O}_{le\bar d\bar q}$ and ${\cal O}_{le\bar e'\bar l'}$
(see  right-hand side of   Fig.~\ref{feynman2loop}),
 we get respectively
\be
\frac{d}{d\ln \mu}
\left(\begin{array}{c} C_{eB}\\ C_{eW}\end{array}\right)
=\frac{y_d g^3}{(16\pi^2)^2}\frac{N_c}{4}
\left(
\begin{array}{c} 
3 t_{\theta_W} Y_Q + 4 t_{\theta_W}^3 (Y_L+Y_e)(Y_Q^2+Y_d^2) \\  
\frac{1}{2} + 2 t_{\theta_W}^2 (Y_L+Y_e)Y_Q
\end{array}\right)
C_{le\bar d\bar q}
\label{r4}
\ee
and
\be
\frac{d}{d\ln \mu}
\left(\begin{array}{c} C_{eB}\\ C_{eW}\end{array}\right)
=\frac{y_{e'} g^3}{(16\pi^2)^2}\frac{1}{4}
\left(
\begin{array}{c} 
3 t_{\theta_W} Y_L+ 4 t_{\theta_W}^3 (Y_L+Y_e)(Y_L^2+Y_e^2) \\  
\frac{1}{2} + 2 t_{\theta_W}^2 (Y_L+Y_e)Y_L
\end{array}\right)
C_{le\bar e'\bar l'}\,.
\label{r5}
\ee

We  summarize our results  by schematically presenting in Fig.~\ref{fig:2-loop_running}
the mixing patterns of the effective operators contributing to the electron dipoles at 2-loop order.
For completeness we also include the 2-loop mixing of the ${\cal O}_{\widetilde W}$
operator, which, at 1-loop order, only induces finite corrections to the electron dipoles.
We also provide below the leading-log approximation to the electron EDM 
that can be good enough
since the new physics scale is not constrained yet to be   far away from the electroweak scale,
and therefore we do not need to resum the logs by exactly solving the RGEs.
We find that the one-loop corrections are
\begin{equation}
\displaystyle  \frac{d_e}{e} \simeq - \frac{1}{16 \pi^2}\frac{\sqrt{2}v}{\Lambda^2}\left[
y_u\, {\rm Im}\, C_{luqe}
- y_e \left(2 \tilde{\kappa}_{\gamma\gamma}
 + \frac{1 - 4 s_{\theta_W}^2}{s_{\theta_W} c_{\theta_W}}
  \tilde{\kappa}_{\gamma Z} - \frac{\Lambda^2}{2 v^2}\delta\tilde{\kappa}_\gamma\right)\right] \ln\frac{\Lambda^2}{m_h^2}\,.
\label{deluqe}
\end{equation}
The double-log 2-loop corrections are given by
\begin{eqnarray}
\displaystyle  \frac{d_e}{e} & \displaystyle \simeq \frac{1}{(16 \pi^2)^2}\frac{\sqrt{2}v}{\Lambda^2} \frac{1}{8} {\rm Im}\Bigg[\!\!&
\displaystyle g y_e \Big(y_{e'} (11 + 9 t_{\theta_W}^2) c_{e'W} + 15 y_u (3 + t_{\theta_W}^2) C_{uW}
+ 3 y_d (3 + t_{\theta_W}^2) C_{dW}\Big)\nonumber\\
&& -\, 3 \frac{g}{t_{\theta_W}} y_e \Big( y_{e'} (1 + 7 t_{\theta_W}^2) c_{e'B} - 3 y_u (1 + 7 t_{\theta_W}^2) C_{uB}
+ y_d (3 + 5 t_{\theta_W}^2) C_{dB}\Big)\nonumber\\
&& +\,2 g^2 y_u (3 + 5 t_{\theta_W}^2) C^{(1)}_{lequ}
\, +\,y_e g^3 \left( 13+3\tan\theta_W^2   \right) C_{\widetilde{W}}
\Bigg] \ln^2 \frac{\Lambda^2}{m_h^2}\,.
\end{eqnarray}
Finally, the single-log 2-loop corrections are
\begin{equation}\label{eq:running_2-loop_single_log}
\displaystyle  \frac{d_e}{e} \simeq -\frac{g^2}{(16 \pi^2)^2}\frac{\sqrt{2}v}{\Lambda^2} {\rm Im}\left[
\frac{3}{8}t_{\theta_W}^2\, C_{y_e}
+ y_d \frac{1}{8}t_{\theta_W}^2 C_{le\bar d\bar q}
+ y_{e'} \frac{1}{8}\left(2 + 9 t_{\theta_W}^2\right) C_{le\bar e'\bar l'}
\right] \ln\frac{\Lambda^2}{m_h^2}\,.
\end{equation}

Notice that we did not include in the above formula finite corrections that can arise at the matching scales.
In fact, few  other dimension-6 operators can enter in this way into the renormalization of the electron EDM.
As we saw in Sec.~\ref{sec:one-loop}, an example is provided by the ${\cal O}_{\widetilde W}$ operator, whose
dominant corrections to the electron EDM arise as finite one-loop contributions. An analogous result is valid
for the $H^3 \bar f f$ operator ($f=u,d,e'$)   that modify the quarks and heavy lepton Yukawa couplings. 
These operators start to contribute
to the electron EDM at the two-loop level through finite Barr--Zee-type diagrams~\cite{Barr:1990vd}. We will discuss these operators in Sec.~\ref{sec:cp-violating-yukawas}.

\subsection{Dimension-8 operators}

Since we are calculating corrections to the EDM at the two-loop level coming from dimension-6 operators,
it is appropriate to ask whether dimension-8 operators can also give similar contributions.
The effects of these operators are  suppressed with respect to dimension-6 operators by an extra $m_W^2/\Lambda^2$, but this could be overcome if their contributions to the EDM arises at the one-loop level instead of two-loops.
We will only be interested here in dimension-8 operators that can be generated from integrating new physics at tree-level
and that have not been constrained  from previous dimension-6 operators. For example,
dimension-8 operators involving  extra $|H|^2$ or extra derivatives will not be relevant.  
We only find two of these operators,
\be
{\cal O}_{ldqe}=(\bar L_L d_RH)(\bar Q_L e_RH)
\ ,\ \ \ \ \ 
{\cal O}_{le'l'e}=(\bar L_L e_R'H)(\bar L_L' e_RH)\,,
\ee
that at the one-loop level can mix with  dimension-8 operators of dipole type,
\be
 {\cal O}_{h^2eW}=|H|^2 {\cal O}_{eW}
 \ ,\ \ \ \ \ 
 {\cal O}_{h^2eB}=|H|^2 {\cal O}_{eB}\,,
\ee
similarly to \eq{eq:cyuyetoWdipolerge}: 
\be
\frac{d}{d\ln \mu}
\left(\begin{array}{c} C_{h^2eB}\\ C_{h^2eW}\end{array}\right)
=\frac{y_dg}{16\pi^2}
\left(
\begin{array}{c} -\frac{1}{2}  t_{\theta_W} N_c (Y_Q+Y_d) \\
\frac{1}{4} N_c
\end{array}\right)
C_{ldqe}\,,
\label{eq:cyuyetoWdipolerge8}
\ee
and equivalently for $C_{le'l'e}$ with the replacement $y_d\to y_e$, $Y_Q\to Y_L$ and $Y_d\to Y_e$.
We get a  contribution to EDM of order
\be
\frac{d_e}{e}\simeq 
-\frac{\sqrt{2}v^3}{\Lambda^4}\frac{y_d {\rm Im} [C_{ldqe}]}{16\pi^2}\frac{N_c}{24}\ln\frac{\Lambda^2}{m_W^2}\,.
\label{total8}
\ee

\subsection{Threshold effects at the EW scale: the impact of CP-violating Yukawa's}\label{sec:cp-violating-yukawas}

The log-enhanced contributions we considered in the previous sections are expected to give the dominant
corrections to the electron EDM. 
Nevertheless, being this logarithm not always so large,
there are cases in  which finite corrections can be more important.
A  noticeable example, that we will discuss here, is 
the case of  two-loop corrections to the EDM 
generated  when integrating out the top and the Higgs at the EW scale.
These corrections come from CP-violating Yukawa couplings induced by the SM dimension-6 operator ${\cal O}_{y_f}$ at the EW scale:
\be
c_f h \bar f_Lf_R\ ,\ \ \ \
c_f=\frac{v^2}{\sqrt{2}\Lambda^2}C_{y_f}\,.
\label{cpviolatingy}
\ee
For the particular case of $C_{y_e}$  we obtain,  from  Barr--Zee diagrams,  that these contributions are given by
\begin{equation}\label{eq:Oye_finite}
\frac{d_e}{e} \simeq - \frac{16}{3 \sqrt{2}} \frac{e^2}{(16 \pi^2)^2} v \left(2 + \ln\frac{m_t^2}{m_h^2}\right)
\frac{{\rm Im}\, C_{y_e}}{\Lambda^2}\,,
\end{equation}
where we only kept the leading corrections due to diagrams with a virtual photon and a top loop. Contributions from
a  virtual $Z$-boson are highly suppressed by the small vector $Z$ coupling to the electron, whereas  diagrams involving
light quarks are suppressed by the light quark Yukawa's. Diagrams involving a virtual $W$ boson are expected to be subleading with respect to the photon contribution, although not negligible. For simplicity we however neglect these contribution as often done in the literature. Notice that in the above formula we only included the leading terms
in an expansion for large $m_t^2/m_h^2$, which reproduce the full result with an accuracy $\sim 10\%$.\footnote{We
report the full expression in Appendix~\ref{app:Barr--Zee}.}
The appearance of  $\ln m_t^2/m_h^2$ in \eq{eq:Oye_finite} can be understood as a RG running  of the electron EDM from
the top mass, where the top is integrated out generating a $hF^{\mu\nu}F_{\mu\nu}$ term at the one-loop level, down to the Higgs mass. In particular, this arises from a loop diagram involving $hF^{\mu\nu}F_{\mu\nu}$ and the CP-violating Yukawa \eq{cpviolatingy}.

Comparing the result in \eq{eq:Oye_finite} with the two-loop running in \eq{eq:running_2-loop_single_log},
we find that the former is typically dominant and is a factor of few larger than the latter for
a cut-off scale in the $10\;$TeV range. 
This is because the contributions
of \eq{eq:running_2-loop_single_log}  are slightly smaller than the naive estimate since they are proportional to $g'$ and are suppressed by an accidental factor $3/8$.
The two contributions become  comparable only for a very large cut-off scale $\sim 10^4\;$TeV.

Similarly,  $C_{y_{u,d}}$ and $C_{y_{e'}}$,
 which can give CP-violating corrections to the quark and to the heavy lepton Yukawa's,
can lead  to  finite Barr--Zee contributions to the electron EDM. For the top  case, we have
\begin{equation}\label{eq:Oyt_finite}
\frac{d_e}{e} \simeq - \frac{e^2}{(16 \pi^2)^2} 2 \sqrt{2} N_c Q_t^2 \frac{m_e}{m_t} v \left(2 + \ln\frac{m_t^2}{m_h^2}\right)
\frac{{\rm Im}\, C_{y_t}}{\Lambda^2}\,.
\end{equation}
Again, the $\ln m_t^2/m_h^2$ can be understood as a RG running of the electron EDM from the top mass, where a $hF\tilde F$
term is generated from the CP-violating top Yukawa after integrating out the top, down to the Higgs mass.

\subsection{EFT below the electroweak scale and relevant RGEs}

Let us now discuss the RG running effects below the EW scale.
At the scale $\mu\sim m_W$, we must integrate out all the heavy SM particles, the $W$, $Z$, the Higgs and top,
and work for $\mu<m_W$ with the EFT built with the light fermions, the photon and the gluons.
The EDM of the electron arises in this EFT from a dimension-5 operator
\be
{\cal O}_{e\gamma}=\bar e_L \sigma^{\mu\nu} e_RF_{\mu\nu}\,,
\ee
whose coefficient, from the tree-level matching   with the SM EFT at the EW scale, is given by
\be
C_{e\gamma}(m_W)=
\frac{v}{\sqrt{2}\Lambda}\left(s_{\theta_W}\, C_{eW}(m_W)-c_{\theta_W}\, C_{eB}(m_W)\right)
+\frac{v^3}{2\sqrt{2}\Lambda^3}\left(s_{\theta_W}\, C_{h^2eW}(m_W)-c_{\theta_W}\,C_{h^2eB}(m_W)\right)\,.
\ee
Other relevant operators below the EW scale are four-fermion operators made with the light SM fermions.
The matching at the EW scale is given by
\bea
C_{luqe}(m_W){\cal O}_{luqe}&=&C_{euue}(m_W) (\bar e_L u_R)(\bar u_L e_R)+\cdots\,,\nonumber\\
C^{(1)}_{lequ}(m_W){\cal O}^{(1)}_{lequ}&=&C_{eeuu}(m_W) (\bar e_L e_R)(\bar u_L u_R)+\cdots\,,\nonumber\\
C_{le\bar d\bar q}(m_W){\cal O}_{le\bar d\bar q}&=&C_{ee\bar d\bar d}(m_W) (\bar e_L e_R)(\bar d_R d_L)+\cdots\,,\nonumber\\
C_{le\bar e\bar l}(m_W){\cal O}_{le\bar e\bar l}&=&C_{le\bar e\bar e}(m_W) (\bar e_L e_R)(\bar e'_R e'_L)+\cdots\,,
\label{mat1}
\eea
and 
\bea
C_{ldqe}(m_W){\cal O}_{ldqe}&=&\frac{v^2}{2\Lambda^2}C_{edde}(m_W) (\bar e_L d_R)(\bar d_L e_R)+\cdots\,,\nonumber\\
C_{le'l'e}(m_W){\cal O}_{le'l'e}&=&\frac{v^2}{2\Lambda^2}C_{ee'e'e}(m_W) (\bar e_L e'_R)(\bar e'_L e_R)+\cdots\,,\nonumber\\
C_{ldqe}(m_W){\cal O}_{leqd}&=&\frac{v^2}{2\Lambda^2}C_{eedd}(m_W) (\bar e_L e_R)(\bar d_L d_R)+\cdots\,,\nonumber\\
C_{le'l'e}(m_W){\cal O}_{lel'e'}&=&\frac{v^2}{2\Lambda^2}C_{eee'e'}(m_W) (\bar e_L e_R)(\bar e'_L e'_R)+\cdots\,,
\label{mat2}
\eea
where we have also included the matching of the dimension-8 SM operators 
\be
{\cal O}_{leqd}=(\bar L_L e_RH)(\bar Q_L d_RH)
\ ,\ \ \ \ \ 
{\cal O}_{lel'e'}=(\bar L_L e_RH)(\bar L_L' e'_RH)\,.
\ee
The above four-fermion operators can enter into the anomalous dimension of ${\cal O}_{e\gamma}$ at the one or two
loop level. Using our previous results, 
we can easily extract the RGE for $C_{e\gamma}$  since the renormalization  of the photon
 can be read from that of the ${\rm U}(1)_Y$-boson in the SM. 
At the one-loop level we only have, equivalently to \eq{eq:cyuyetoWdipolerge} and \eq{eq:cyuyetoWdipolerge8}, (see also \cite{Jenkins:2017dyc})
\be
\frac{d}{d\ln \mu} C_{e\gamma}
=-\frac{e}{16\pi^2}\frac{1}{\sqrt{2}}\left(\frac{m_u}{\Lambda}
  N_c Q_u
C_{euue}
+\frac{m_dv^2}{2\Lambda^3}
  N_c Q_d
C_{edde}
+\frac{m_{e'}v^2}{2\Lambda^3}
   Q_eC_{ee'e'e}
\right)
\,,
\label{r1}
\ee
where $Q_f$ refers to the EM charge  of the fermion $f$.
At the two-loop level, we can have contribution to $C_{e\gamma}$ either
by a one-loop mixing of ${\cal O}_{eeff}$ with ${\cal O}_{effe}$ ($f=u,d,e'$),
similarly to \eq{ltol}, 
\be
\frac{d}{d\ln \mu}  C_{effe}=\frac{e^2}{\pi^2}Q_eQ_fC_{eeff}\,,
\label{r1b}
\ee
 or by a direct contribution to $C_{e\gamma}$:
\be
\frac{d}{d\ln \mu}
C_{e\gamma}
=\frac{e^3}{(16\pi^2)^2}\left(\frac{m_d}{\Lambda}4 N_c  Q_eQ^2_d
 C_{ee\bar d\bar d}+
\frac{m_e}{\Lambda}4 
 Q_e^3 
C_{ee'\bar e'\bar e}\right)\,.
\label{r2d}
\ee
The RGE running  should be considered from $m_W$ down to the mass of the heaviest fermion in the four-fermion
operator, where the state should be integrated out.\footnote{We have also to include here  self-renormalization
effects. These  are however small and generically correct the bounds on the Wilson coefficients
by roughly $10\%$.}
Therefore this running can be important for lighter fermions.

We can compare our results with those obtained from Barr-Zee diagrams arising from CP-violating Yukawa
interactions \eq{cpviolatingy}.
For  light quarks and leptons $e'$, one finds respectively
\begin{equation}\label{eq:Oyqlight_finite}
\frac{d_e}{e} \simeq - \frac{e^2}{(16 \pi^2)^2} 4 N_c Q_q^2 v \frac{m_e m_q}{m_h^2} \left(\ln^2\frac{m_q^2}{m_h^2}
+ \frac{\pi^2}{3}\right) \frac{{\rm Im}\, C_{y_q}}{\Lambda^2}\,,
\end{equation}
and
\begin{equation}\label{eq:Oye'_finite}
\frac{d_e}{e} \simeq - \frac{e^2}{(16 \pi^2)^2} 4 Q_e^2 v \frac{m_e m_{e'}}{m_h^2} \left(\ln^2\frac{m_{e'}^2}{m_h^2}
+ \frac{\pi^2}{3}\right) \frac{{\rm Im}\, C_{y_{e'}}}{\Lambda^2}\,.
\end{equation}
The double logarithm in Eqs.~(\ref{eq:Oyqlight_finite}) and (\ref{eq:Oye'_finite}) can be understood from an EFT perspective as 
the  RG  running  from the Higgs mass, where integrating the   Higgs   generates  a complex $C_{eeff}$ from \eq{cpviolatingy},  down to the light fermion masses. The relevant RGEs are indeed those  in \eq{r1} and \eq{r1b}.

For light quarks, however, 
a better bound on the corresponding Wilson coefficient can be obtained from constraints
on  CP-violating electron-nucleon interactions, that the   ACME collaboration has also recently  reported \cite{ACME}:
\be
-\frac{G_F}{\sqrt{2}}i\bar e\gamma_5 e \bar N\left(C^{(0)}_S+C^{(1)}_S\sigma_3\right)N
\ ,\ \ \ \ 
C_S=C_S^{(0)}+\frac{Z-N}{Z+N}C_S^{(1)}<7.3\cdot 10^{-10}\,.
\ee
Neglecting isospin breakings that in the ThO are small \cite{ACME}, we have for the down-quarks
\be
C_S\simeq  C_S^{(0)}=-\frac{{\rm Im}[C_{ee\bar d\bar d}]}{\sqrt{2}G_F\Lambda^2}\langle N|\bar d d|N\rangle\,.
\ee
We have checked that bounds from $d_e$ are slightly better than those from $C_S$
for operators involving the bottom (using $\langle N|\bar bb|N\rangle\simeq 74$ MeV$/m_b$  
\cite{Junnarkar:2013ac}),
 while the bound from   $C_S$ is better
for ligher quarks.

\section{Impact  on  BSM}

\subsection{Power counting of the Wilson coefficients}

So far we have presented the  leading contributions of  the dimension-6 operators
 to the anomalous dimension of the electron  electric dipole operators up to two loops.
All the possible contributing operators are given in Table~\ref{operatortable}.
Nevertheless, the importance of the different operators  depends on the size of 
their Wilson coefficients, which crucially depends on the BSM dynamics.

In the following we will be interested in BSM theories that can be described as weakly-coupled renormalizable theories.
These includes all possible extensions of the SM with extra particles with renormalizable interactions.
In these BSM theories we can classify the  Wilson coefficients as those that can be  
generated at  tree-level and those generated at most at the loop level.
For example, among the operators of Table~\ref{operatortable}, the only ones with Wilson coefficients that can be induced at tree-level by integrating out new heavy states are all the four-fermion operators and ${\cal O}_{y_e}$;
the rest, involving always field-strengths, can only be generated by loops. Therefore, we expect
\be
C_{fV}\sim \frac{g_*^3g}{16\pi^2}\ , \ \ \ C_{V\widetilde V}\sim \frac{g^2_*g^2}{16\pi^2}\ , \ \ \ 
C_{luqe},C_{lequ},C_{le\bar d\bar q},C_{le\bar e'\bar l'},C_{y_e}\sim g^2_*\,,
\ee
where $g_*$ refers to  a generic coupling of the BSM dynamics to the SM, $f=e,u,d$ and $V=W,B$.

\begin{table}[t]
\begin{center}
\begin{tabular}{|c|c|}
\hline 
${\cal O}_{luqe}$ 
& Scalar $\bf (3,2,7/6)$\\
& Scalar $\bf (\bar 3,1,1/3)$\\
\hline 
${\cal O}^{(1)}_{lequ}$  & Scalar $\bf (1,2,1/2)$\\
& Scalar $\bf (\bar 3,1,1/3)$\\
\hline
${\cal O}_{le\bar d \bar q}$ & Vector $\bf (\bar 3,2,5/6)$\\
& Vector $\bf (3,1,2/3)$\\ 
& Scalar $\bf (1,2,1/2)$\\
\hline
${\cal O}_{le\bar e' \bar l'}$  & Vector $\bf ( 1,1,0)$\\
& Vector $\bf (1,2,1/2)$\\
& Scalar $\bf ( 1,2,1/2)$\\
\hline
 \end{tabular}
 \hspace{10mm}
 \begin{tabular}{|c|c|}
\hline
${\cal O}_{y_e}$  & Fermion $\bf (1,2,-1/2)\oplus(1,1(3),-1)$\\
& Fermion $\bf (1,2,-1/2)\oplus(1,1(3),0)$\\
& Fermion $\bf (1,2,-3/2)\oplus(1,1(3),0)$\\
& Scalar $\bf (1,2,1/2)$\\
\hline
 \end{tabular}
 \end{center}
\caption{\it States  transforming under the SM group ${\rm SU}(3)_c \times {\rm SU}(2)_L \times {\rm U}(1)_Y$ 
contributing at the tree-level to the  operators of Table~\ref{operatortable}.}
\label{statesattree}
\end{table}

In this class of BSM theories the contributions to the electron EDM have the following  loop expansion.
The  leading contributions
are of order $d_e/e \sim (g_*^2/16\pi^2) (m_u/\Lambda^2) \ln({\Lambda/m_W})$
and can arise from those particular BSM that contribute to ${\cal O}_{luqe}$ at tree-level. 
There are only two types of particles that can generate
${\cal O}_{luqe}$ at tree-level (see Table~\ref{statesattree}), the leptoquarks $R_2$ and $S_1$ that will be discussed below.

Next,  we can have  BSM dynamics contributing directly to the Wilson coefficients of the electron dipole operators, that can give corrections of one-loop order (but without a log-enhancement), 
 $d_e/e\sim  (g_*^2/16\pi^2)(g_*v/\Lambda^2) $. 
This can happen in BSM theories that contain  fermions and bosons coupled to the electron, with at least one of them charged, as for example,  the selectron and wino in supersymmetric theories.

Contributing at the two-loop level,  we have those BSM theories inducing  ${\cal O}^{(1)}_{lequ}$
at tree-level, which leads to  a double-log enhanced EDM,  $d_e/e\sim  g^2 g_*^2/(16\pi^2)^2(m_u/\Lambda^2) \ln^2({\Lambda/m_W})$. 
As shown in Table~\ref{statesattree}, this includes BSM theories with extra Higgses.  

Single-log two-loop contributions can come from BSM scenarios generating 
${\cal O}_{V\tilde V}$ at the loop level (those BSM containing extra charged fermions coupled to the Higgs),
or  generating ${\cal O}_{le\bar d\bar q}$, ${\cal O}_{le\bar e'\bar l'}$ or ${\cal O}_{y_e}$ at tree-level
(see  Table~\ref{statesattree}). Also  BSM theories generating ${\cal O}_{luqe}$ at one-loop level
can lead to a single-log two-loop contribution to the electron EDM, as we will see later for the case of the MSSM.

Finally,  BSM theories with extra  ${\rm SU(2)}_L$ fermions can generate  ${\cal O}_{\widetilde W}$ at the loop-level, leading to a
contribution to the electron EDM    at the two-loop level with no log enhancement. On the other hand,
BSM theories contributing to the EDM of a fermion different from the electron are expected to give negligible effects to the electron EDM, as these arise at best at the three-loop level.

Examples of these classes of BSM  theories will be given below.
We must also notice that there is a large class of low-energy effective descriptions of strongly-coupled theories that follow
the same power counting described above. These are those theories that were assumed to follow the   "minimal coupling" assumption  \cite{Liu:2016idz}, and correspond to holographic models as well as their deconstructed versions.

It is also important  to  keep in mind  that   operators
of Table~\ref{operatortable}   containing the fields  $\bar L_L$ and $e_R$
 can potentially give a contribution to the electron mass.
This places a constraint on the natural size of their Wilson coefficients. In particular, we find
\be\label{eq:natural_size}
\left\{\frac{C_{eV}\, v}{16\pi^2}\ ,\ 
\frac{C_{y_e}\, v^3}{\Lambda^2}\ , \  
\frac{C_{lequ}\, m_u}{16\pi^2}\ ,\ 
\frac{C_{luqe}\, m_u}{16\pi^2}\ ,\   
\frac{C_{le\bar d \bar q}\, m_d}{16\pi^2}\ ,  \ 
\frac{C_{le\bar e \bar l}\, m_{e'}}{16\pi^2}\right\}\lesssim m_e\,.
\ee
In fact, in most of the UV-complete BSM theories
(e.g. supersymmetry, composite Higgs or theories with flavor symmetries only broken by Yukawas)
we expect operators with chirality flips to carry  yukawa couplings, i.e.,
\be
C_{fV}\propto y_f\ ,\ \ \ 
C_{y_e}\propto y_e\ ,\ \ \  
C_{lequ}\propto y_e y_u\ ,\ \ \
C_{luqe}\propto y_e y_u\ ,\ \ \
C_{le\bar d \bar q}\propto y_e y_d\ ,\ \ \
C_{le\bar e' \bar l'}\propto y_e y_{e'}\,,
\label{ndacoeff}
\ee
implying that we only expect sizable contributions to the electron EDM from
four-fermion operators involving the third family. All these considerations can be useful 
for a proper interpretation of  the recent ACME bound.

\renewcommand{\arraystretch}{1.3}
\begin{table}[t]
\begin{center}
\begin{tabular}{|c|c|}
\multicolumn{2}{c}{tree-level}\\
\hline
$C_{eW}$&  $5.5 \times 10^{-5}\, y_eg$\\
$C_{eB}$ & $5.5 \times 10^{-5}\, y_eg'$\\
  \hline
\multicolumn{2}{c}{\rule[-7pt]{0pt}{2em}one-loop}\\
\hline
   $C_{luqe}$ & $1.0 \times 10^{-3}\, y_ey_t$\\
   $C_{W\widetilde W}$ & $4.7 \times 10^{-3}\, g^2$\\
$C_{B\widetilde B}$ & $5.2 \times 10^{-3}\, g^{\prime\, 2}$\\
$C_{W\widetilde B}$ & $2.4 \times 10^{-3}\, gg'$\\
$C_{\widetilde W}$& $6.4 \times 10^{-2}\, g^3$\\
\hline
  \end{tabular}\hspace{10mm}
  \begin{tabular}{|c|c|}
  \multicolumn{2}{c}{two-loops}\\
  \hline
   $C_{lequ}$& $3.8 \times 10^{-2}\, y_ey_t$\\
   $C_{\tau W}$& $260\, y_\tau g$\\
   $C_{\tau B}$& $380\, y_\tau g'$\\
   $C_{tW}$& $6.9 \times 10^{-3}\, y_t g$\\
   $C_{tB}$& $1.2 \times 10^{-2}\, y_t g'$\\
   $C_{bW}$& $64\, y_b g$\\
   $C_{bB}$& $47\, y_b g'$\\
   $C_{le \bar d\bar q}$& $10\, y_e y_t (y_t/y_b)$\\
   $C_{le\bar e'\bar l'}$& $0.63\, y_ey_t (y_t/y_\tau)$\\
   \hline
  \end{tabular}\hspace{10mm}
  \begin{tabular}{|c|c|}
  \multicolumn{2}{c}{two-loops finite}\\
  \hline
   $C_{y_e}$& $20\, y_e \lambda_h$\\
   $C_{y_t}$& $20\, y_t \lambda_h$\\
   $C_{y_b}$& $4.1 \times 10^3 \, y_b \lambda_h$\\
   $C_{y_\tau}$& $4.8 \times 10^3\, y_\tau \lambda_h$\\
   \hline
  \end{tabular}
\caption{\it Bounds on the Wilson coefficients coming from \eq{expbound}
taking $\Lambda=10$ TeV.
For a better appreciation of the bound, we have extracted the Yukawa,
gauge or Higgs coupling ($\lambda_h=0.1$) that
we naturally expect to carry these Wilson coefficients. For   $C_{le \bar
d\bar q}$ and  $C_{le\bar e'\bar l'}$
we have further extracted a factor $(y_t/y_b)$ and  $(y_t/y_\tau)$
respectively to reflect the fact that these coefficients
can be potentially larger  consistently with their  natural sizes \eq{eq:natural_size}.  }
\label{operatortable_bounds}
\end{center}
\end{table}

In Table~\ref{operatortable_bounds} we list the bounds on individual Wilson coefficients that can be inferred from the
new electron EDM measurement \eq{expbound}. To derive the bounds we considered the various Wilson
coefficients one-by-one. Although typical BSM theories give rise to simultaneous contributions to several
Wilson coefficients, strong cancellations are typically not present. In such situation the bounds obtained on
single Wilson coefficients remain approximately valid.\footnote{Bounds on effective operators coming from measurements
of the electron EDM were previously derived in the literature in refs.~\cite{Cirigliano:2016njn,Frigerio:2018uwx}.}

\subsection{Leptoquarks and extra Higgs}

As a first example of an application of the above EFT analysis, 
we focus here on new-physics models containing states of Table~\ref{statesattree}.
In particular, we focus on leptoquarks and heavy Higgs-like states.\footnote{Notice that, in addition to the electron EDM,
leptoquarks and heavy Higgs-like states, as well as supersymmetric scenarios, can also be constrained
by the EDM of $^{199}$Hg atom through the CP-odd electron-nucleon interaction~\cite{Yamanaka:2017mef}.}

As can be seen from Table~\ref{statesattree},
four leptoquark multiplets can give rise to electron EDM contributions up to two-loop order.\footnote{See Ref.~\cite{Dorsner:2016wpm} for a review of leptoquark properties and for the nomenclature. See also Ref.~\cite{Dekens:2018bci} for a recent reappraisal of the contributions of the
scalar leptoquarks to electron and light quark EDMs.}
Among scalar leptoquarks only the $R_2$
and the $S_1$ multiplets
give rise to contributions to $d_e$. In both cases the contributions arise at one-loop level and include a logarithmically-enhanced term.
On the other hand, vector leptoquarks, in particular the $V_2$
and the $U_1$ multiplets
can contribute to the electron EDM at one-loop order only with finite contributions.

We analyze the various cases in the following, providing the matching with the EFT operators in the limit of heavy
multiplet masses. For the scalar leptoquarks, we also compare the leading running contributions to the EDM
with the full results, which are already known in the literature. For simplicity we only include couplings to third-generation quarks,
since interactions with the light generations give rise to EDM contributions suppressed by the light-fermion Yukawa couplings.

\subsubsection{Scalar leptoquarks}

We start our discussion with the case of scalar leptoquarks.

\paragraph{The $R_2$ leptoquark}

The first case we consider is the $R_2$ multiplet, whose ${\rm SU}(3)_c \times {\rm SU}(2)_L \times {\rm U}(1)_Y$ quantum
numbers are $({\bf 3}, {\bf 2}, {\bf 7/6})$. The Lagrangian describing the relevant leptoquark interactions with the SM fermions is
\begin{equation}
{\cal L} = - y_2^{RL} \overline t_R R^a \varepsilon^{ab} L^b_{L_1} + y_2^{LR} \overline e_R R^{a*} Q^a_{L_3} + {\rm h.c.}\,,
\end{equation}
where $L_{L_i}$ and $Q_{L_i}$ labels the $i$-generation lepton and quark respectively.
In the limit of large mass, the $R_2$ leptoquark gives rise to a contribution to the ${\cal O}_{l u q e}$ effective operator,
namely
\begin{equation}
{\cal L}_{\it eff}^{R_2} = \frac{y_2^{LR*}y_2^{RL*}}{m_{R_2}^2} {\cal O}_{l u q e} + {\rm h.c.}\,.
\end{equation}
Using  \eq{deluqe}, we can obtain the  log-enhanced one-loop contribution to the electron EDM:
\begin{equation}\label{eq:R_2_EDM_app}
\frac{d_e}{e} \simeq \frac{1}{8 \pi^2} \frac{m_t}{m_{R_2}^2} {\rm Im}\left(y_2^{LR} y_2^{RL}\right) \ln \frac{m_{R_2}^2}{m_t^2}\,.
\end{equation}

The full one-loop contribution to the electron EDM is also known in the literature\cite{Fuyuto:2018scm}
\begin{equation}\label{}
\frac{d_e}{e} = \frac{3}{32 \pi^2} \frac{m_t}{m_{R_2}^2} {\rm Im}\left(y_2^{LR} y_2^{RL}\right)
\left[Q_t I_2\left(m_t^2/m_{R_2}^2\right) + Q_{LQ} J_2\left(m_t^2/m_{R_2}^2\right)\right]\,,
\end{equation}
where $Q_t = 2/3$ and $Q_{LQ} = 5/3$ are the electric charges of the top quark and $R_2$ leptoquark, while the
$I_2$ and $J_2$ functions are given by
\begin{equation}\label{eq:R_2_EDM_full}
I_2(x) = \frac{1}{(1-x)^3}(-3+4x-x^2-2\ln x)\,,
\qquad
J_2(x) = \frac{1}{(1-x)^3}(1-x^2+2x\ln x)\,.
\end{equation}
One can easily check that the leading logarithmic term in Eq.~(\ref{eq:R_2_EDM_full}) agrees with the result of the
EFT calculation in Eq.~(\ref{eq:R_2_EDM_app}). In fact,
the leading-log contribution in Eq.~(\ref{eq:R_2_EDM_app}) provides a quite good approximation of the full result
even for relatively small leptoquark masses. The discrepancy is below $25\%$ for $m_{R_2} > 300\;{\rm GeV}$ and below
$10\%$ for $m_{R_2} > 360\;{\rm GeV}$.

The recent  bound on the electron EDM  \eq{expbound} translates into the  constraint\footnote{Notice that \eq{eq:bound_R2}, as well as
Eqs.~(\ref{eq:bound_S1}), (\ref{eq:bound_V2}), (\ref{eq:bound_U1}), (\ref{eq:bound_H1}), (\ref{eq:bound_H2}), (\ref{eq:bound_H3}) and (\ref{eq:bound_H4}), are valid as far as the mass bound is $\gtrsim 1\;$TeV. Below this value the leading-log
approximation is not accurate since threshold effects can become relevant.}
\be\label{eq:bound_R2}
m_{R_2} \gtrsim 420\;{\rm TeV}\, \sqrt{\frac{|{\rm Im}(y_2^{LR}y_2^{RL})|}{y_ey_t} \left(1 + 0.075 \ln \frac{|{\rm Im}(y_2^{LR}y_2^{RL})|}{y_ey_t}\right)}\,,
\ee
where we have normalized  $y_2^{LR}y_2^{RL}$ to  the electron and top Yukawa coupling, following
 the estimates presented in  \eq{ndacoeff}.
 
\paragraph{The $S_1$ leptoquark}

The second leptoquark state that can give rise to one-loop contributions to the electron EDM is the $S_1$ multiplet,
which has $({\bf \overline 3}, {\bf 1}, {\bf 1/3})$ quantum numbers. Its interactions with the SM fermions can be parametrized by
\begin{equation}
{\cal L} = y_1^{LL} \overline Q_{L_3}^{C\,a} S_1 \varepsilon^{ab} L_{L_1}^b + y_1^{RR} \overline t_R S_1 e_R + {\rm h.c.}\,,
\end{equation}
where the $C$ superscript denotes the charge conjugation operation, namely $\psi^C \equiv C \overline \psi^T$
with $C = i \gamma^2 \gamma^0$.
Integrating out  $S_1$ gives rise to a contribution to  ${\cal O}_{l u q e}$ and ${\cal O}^{(1)}_{\ell e q u}$, namely
\begin{equation}
{\cal L}_{\it eff}^{S_1} = \frac{y_1^{LL*}y_1^{RR}}{m_{S_1}^2} \left[{\cal O}_{l u q e} + {\cal O}^{(1)}_{l e q u}\right] + {\rm h.c.}\,.
\end{equation}
Therefore, from \eq{deluqe}, we obtain the following   log-enhanced one-loop contribution to the electron EDM 
\begin{equation}\label{eq:S_1_EDM_app}
\frac{d_e}{e} \simeq \frac{1}{8 \pi^2} \frac{m_t}{m_{S_1}^2} {\rm Im}\left(y_1^{LL} y_1^{RR*}\right) \ln \frac{m_{S_1}^2}{m_t^2}\,.
\end{equation}

The full one-loop result reads~\cite{Dorsner:2016wpm}
\begin{equation}\label{eq:S_1_EDM_full}
\frac{d_e}{e} = \frac{1}{32 \pi^2} \frac{m_t}{m_{S_1}^2} {\rm Im}\left(y_1^{LL} y_1^{RR*}\right) G\left({m_t^2}/{m_{S_1}^2}\right)\,,
\end{equation}
where the $G$ function is defined by
\begin{equation}
G(x) = \frac{1}{(1-x)^3} \left(5 - 8 x + 3 x^2 +2 (2 - x) \ln x\right)\,.
\end{equation}
We find that the leading-log approximation in Eq.~(\ref{eq:S_1_EDM_app}) is in fair agreement with the full result, the difference being $\lesssim 30\%$ for $m_{S_1} \gtrsim 220\;{\rm GeV}$.

\eq{expbound}  translates into the  bound
\be\label{eq:bound_S1}
m_{S_1} \gtrsim 400\;{\rm TeV}\, \sqrt{\frac{|{\rm Im}(y_1^{LL} y_1^{RR*})|}{y_ey_t} \left(1 + 0.081 \ln \frac{|{\rm Im} (y_1^{LL} y_1^{RR*})|}{y_ey_t}\right)}\,.
\ee

\subsubsection{Vector leptoquarks}
  
We now consider the case of vector leptoquarks. Before specializing the discussion to the $V_2$ and $U_1$ cases,
we discuss some generic features of these models.

As we already mentioned, vector leptoquarks can give rise to a finite contribution to the electron EDM at the $1$-loop order.
In order to compute the EDM effects, one needs first of all to embed the vector leptoquark
into a well-behaved (i.e.~renormalizable) UV theory.\footnote{If the vector leptoquarks are described through
the non-renormalizable Proca Lagrangian, divergent contributions are obtained for the EDM~\cite{Biggio:2016wyy}.}
For this purpose one can consider GUT-like extensions of the SM, as done in refs.~\cite{Lavoura:2003xp,Biggio:2016wyy}, which always lead to the following Lagrangian for the couplings of the Leptoquark $V_\mu$ to the photon $A_\mu$
\begin{equation}
{\cal L} \supset - i\,e\,Q_V\left[(\partial_\mu V_\nu^\dagger - \partial_\nu V_\mu^\dagger) A^\mu V^\nu
- (\partial_\mu V_\nu - \partial_\nu V_\mu) A^\mu V^{\nu\dagger}
+ (V_\mu V_\nu^\dagger - V_\nu V_\mu^\dagger) \partial^\mu A^\nu\right]\,,
\end{equation}
where $Q_V$ is the leptoquark electric charge.
The couplings of a vector leptoquark to the electron and a quark $q$ can be parametrized as
\begin{equation}
\mathcal{L} \supset g_L \bar e_L \gamma^\mu q_L V_\mu + g_R \bar e_R \gamma^\mu q_R V_\mu + {\rm h.c.}\,.
\end{equation}
The $1$-loop contribution to the electron EDM is given by
\begin{equation}\label{eq:V_leptoquark_edm}
\frac{d_e}{e} = \frac{N_c(Q_V - Q_q)}{8\pi^2}\frac{m_q}{m_V^2}\text{Im}(g_L g_R^\star)\,,
\end{equation}
where $m_q$ and $Q_q$ are the mass and the electric charge of the quark, while $m_V$ is
the leptoquark mass. Notice that contributions come from two type of diagrams, one in which the photon
is attached to the quark line and one in which it is attached to the leptoquark line. The two contributions are therefore proportional to $Q_q$ ad $Q_V$ respectively.

Within our EFT description, these contributions must be matched directly into the Wilson coefficients  $C_{eW}$ and $C_{eB}$
at the scale $m_V$.  Notice, however, that the leptoquark gives also rise to effective $4$-fermion interactions of the form
$(\bar L_L^a\gamma_\mu Q_{L\,a})(\bar  b_R\gamma^\mu e_R)$ that after Fierzing leads to  ${\cal O}_{l e \bar d\bar q}$. 
 As we discussed,  the ${\cal O}_{l e \bar d\bar q}$ operator however  do not contribute at $1$-loop
to the electron EDM  but  at the $2$-loop order.\footnote{If the $4$-fermion operators are written in vector-current form and $\overline{\rm MS}$ regularization is used, the $1$-loop contribution to the electron EDM proportional to the quark charge would be matched onto the Wilson coefficient of  $(\bar L_L^a\gamma_\mu Q_{L\,a})(\bar  b_R\gamma^\mu e_R)$. We checked that the finite $1$-loop contribution coming from this operator indeed matches the $Q_q$ term in eq.~(\ref{eq:V_leptoquark_edm}).}

\paragraph{The $V_2$ leptoquark}

Let us now consider $V_2$ multiplet with quantum numbers $({\bf \overline 3}, {\bf 2}, {\bf 5/6})$.
The relevant interactions with the SM fermions read
\begin{equation}
{\cal L} = x_2^{RL} \overline b_R^C \gamma^\mu V^a_{2,\mu} \varepsilon^{ab} L_{L_1}^b + x_2^{LR} \overline Q_{L_3}^{C\,a}
\gamma^\mu \varepsilon^{ab} V^b_{2,\mu} e_R + {\rm h.c.}\,.
\end{equation}
Using the above formulae we find the following $1$-loop contribution to the electron EDM
\begin{equation}
\frac{d_e}{e} = - \frac{5}{8\pi^2}\frac{m_b}{m_{V_2}^2}\text{Im}(x_2^{LR}x_2^{RL\star})\,.
\end{equation}
The new bound from the ACME collaboration leads to
\begin{equation}\label{eq:bound_V2}
m_{V_2} \gtrsim 5.5\,\text{TeV }\, \sqrt{\frac{\text{Im}(x_2^{LR}x_2^{RL\star})}{y_ey_b}}\,.
\end{equation}

\paragraph{The $U_1$ leptoquark}

The second possible vector leptoquark that gives a contributions to the electron EDM is the $U_1$ state, with quantum
numbers $({\bf 3}, {\bf 1}, {\bf 2/3})$. Its Lagrangian reads
\begin{equation}
{\cal L} = x_1^{LL} \overline Q_{L_3}^a \gamma^\mu U_{1,\mu} L^a_{L_1} + x_1^{RR} \overline b_R \gamma^\mu U_{1,\mu} e_R
+ {\rm h.c.}\,.
\end{equation}
The $1$-loop contributions to the electron EDM are given by\,.
\begin{equation}
\frac{d_e}{e} = - \frac{1}{8\pi^2}\frac{m_b}{m_{U_1}^2}\text{Im}(x_2^{RR}x_2^{LL\star})\,.
\end{equation}
The bound in \eq{expbound} leads to
\begin{equation}\label{eq:bound_U1}
m_{U_1} \gtrsim 2.5\,\text{TeV }\, \sqrt{\frac{\text{Im}(x_2^{RR}x_2^{LL\star})}{y_ey_b}}\,.
\end{equation}

\subsubsection{The Heavy Higgs}\label{sec:heavy_higgs}

The last type of massive multiplets we consider are Higgs-like states $H$ 
with quantum numbers $({\bf 1}, {\bf 2}, {\bf 1/2})$.\footnote{We are working in the basis 
at which the SM Higgs has no mass-mixing with the heavy Higgs-like states before EW symmetry breaking. This can always be achieved by an $\SU(2)_L$ rotation.}
Depending on their allowed couplings to the SM fermions they can give rise to different sets of contributions to the
electron EDM.

In general a Higgs-like multiplet can have couplings to all SM fermion species. For simplicity we consider only couplings
to the electron family and to third generation fermions, 
discarding  flavor-violating couplings.
We parametrize the relevant heavy Higgs interactions as\footnote{For simplicity we neglect a possible coupling
$(H^\dagger h)^2 + {\rm h.c.}$, which could give a finite $1$-loop contribution to the  operators ${\cal O}_{V\widetilde V}$ 
that, in turn, give rise to a running for the electron EDM.}
\begin{equation}
{\cal L} = \kappa_e {H}^a \overline L_{L_1}^a e_R + \kappa_t \widetilde {H}^a \overline Q_{L_3}^a t_R
+ \kappa_b {H}^a \overline Q_{L_3}^a b_R + \kappa_\tau {H}^a \overline L_{L_3}^a \tau_R 
+ \eta (H^\dagger h) |h|^2 + {\rm h.c.}\,,
\end{equation}
where $\widetilde {H}^a \equiv \varepsilon^{ab} {H}^{b*}$ and $h$ denotes here the SM Higgs doublet.
In the limit of heavy mass, integrating out the Higgs-like state gives rise to three effective operators that contribute to the electron EDM,
namely
\begin{equation}\label{eq:heavy_Higgs_eff}
{\cal L}_{\it eff}^{H} = - \frac{\kappa_e \kappa_t}{m_{H}^2} {\cal O}^{(1)}_{l e q u}
+ \frac{\kappa_e \kappa_b^*}{m_{H}^2} {\cal O}_{l e \bar d \bar q}
+ \frac{\kappa_e \kappa_\tau^*}{m_{H}^2} {\cal O}_{l e \bar\tau \bar l}
+ \frac{\kappa_e \eta}{m_H^2}{\cal O}_{y_e}\,, 
\end{equation}
where ${\cal O}_{l e \tau l} = (\overline L_{L_1}^a e_R)(\overline \tau_R L_{L_3}^a)$.
As we saw in the previous sections all the effective operators in Eq.~(\ref{eq:heavy_Higgs_eff}) give rise to a contribution
to the electron EDM at two-loop order. There are however some important differences. The ${\cal O}^{(1)}_{l e q u}$
operator  can potentially lead to  the largest contribution, since its effects are enhanced by a double logarithm. The remaining
two operators lead instead to the single logarithmic.

The leading contributions to the electron EDM from the four effective operators are given respectively by
\begin{equation}
\frac{d_e}{e} \simeq \frac{3 g^2 + 5 g'^2}{2 (16 \pi^2)^2} \frac{m_t}{m_H^2} {\rm Im}(\kappa_e \kappa_t) \ln^2 \frac{m_H^2}{m_t^2}\,\,,
\end{equation}
\begin{equation}
\frac{d_e}{e} \simeq -\frac{4}{(16 \pi^2)^2} \frac{m_b}{m_{H}^2} {\rm Im}\left(\kappa_e \kappa_b^*\right)
\left[\frac{e^2}{3} \ln \frac{m_{H}^2}{m_b^2}
+ g^Z_e g^Z_b \ln \frac{m_{H}^2}{m_Z^2}
- \frac{g^2}{16} \ln \frac{m_{H}^2}{m_W^2}\right]\,,
\end{equation}
\begin{equation}
\frac{d_e}{e} \simeq -\frac{4}{(16 \pi^2)^2} \frac{m_\tau}{m_{H}^2} {\rm Im}\left(\kappa_e \kappa_\tau^*\right)
\left[e^2 \ln \frac{m_{H}^2}{m_\tau^2}
+ (g^Z_e)^2 \ln \frac{m_{H}^2}{m_Z^2}
+ \frac{g^2}{16} \ln \frac{m_{H}^2}{m_W^2}\right]\,,
\end{equation}
and
\begin{equation}
\frac{d_e}{e} \simeq -\frac{3 g^2 t_{\theta_W}^2}{8 (16 \pi^2)^2} \frac{\sqrt{2} v}{m_{H}^2} {\rm Im}\left(\kappa_e \eta\right)
\ln \frac{m_{H}^2}{m_h^2}\,.
\end{equation}
These two-loop results agree with the Barr--Zee results \cite{Barr:1990vd,Nakai:2016atk}.

The new bound on the electron EDM leads to the constraints
\be\label{eq:bound_H1}
m_{H} \gtrsim 66\;{\rm TeV}\, \sqrt{\frac{|{\rm Im}(\kappa_e \kappa_t)|}{y_ey_t}} \left(1 + 0.11 \ln \frac{|{\rm Im} (\kappa_e \kappa_t)|}{y_ey_t}\right)\,,
\ee
\be\label{eq:bound_H2}
m_{H} \gtrsim 0.88\;{\rm TeV}\, \sqrt{\frac{|{\rm Im}(\kappa_e \kappa_b)|}{y_e y_t} \left(1 + 0.036 \ln \frac{|{\rm Im} (\kappa_e \kappa_b)|}{y_ey_t}\right)}\,,
\ee
and
\be\label{eq:bound_H3}
m_{H} \gtrsim 1.4\;{\rm TeV}\, \sqrt{\frac{|{\rm Im}(\kappa_e \kappa_\tau)|}{y_ey_t} \left(1 + 0.092 \ln \frac{|{\rm Im} (\kappa_e \kappa_\tau)|}{y_ey_t}\right)}\,,
\ee
for a Higgs coupling to top, bottom and tau respectively. In the presence of a non-vanishing $\eta$,
the contribution to the CP-violating electron Yukawa leads to the additional bound
\be\label{eq:bound_H4}
m_{H} \gtrsim 4.9\;{\rm TeV}\, \sqrt{\frac{|{\rm Im}(\kappa_e \eta)|}{y_e} \left(1 + 0.18 \ln \frac{|{\rm Im} (\kappa_e \eta)|}{y_e}\right)}\,.
\ee
Notice that this bound, for $\eta \sim 1$, is significantly stronger than the ones derived in the presence of couplings to the
bottom or $\tau$, but is much weaker than the one expected in the presence of a sizeable coupling to the top quark.

\subsection{The MSSM}

We will work within the MSSM assuming that the superpartner masses are larger than the EW scale.
CP-violating phases can appear in several terms of the MSSM. Either  in the supersymmetric parameter $\mu$ (that corresponds to the  Higgsino mass), or in    soft supersymmetry breaking terms:
Bino and Wino masses, $M_1$ and $M_2$ respectively, Higgs mixing mass term, $m^2_{12} H_uH_d$, and 
 the scalar trilinears,  e.g. $y_u A_u H_u\widetilde Q_L\tilde u_R$.
Nevertheless,  only  those combinations of MSSM parameters whose phase cannot be removed by redefinitions of fields
can lead to physical CP-violating effects.
A recent analysis of the impact of the new ACME bound on the MSSM can be found in Ref.~\cite{Cesarotti:2018huy}.

The main contribution from the MSSM arise from  one-loop  contributions to $C_{eW}$ and $C_{eB}$, that generate an electron EDM calculated long ago -see for example   \cite{Kizukuri:1992nj}.
 From   Winos and left-handed selectrons ($\tilde L_L$),  we have
\be
\frac{d_e}{e} \simeq  -\frac{g^2}{16\pi^2}\frac{m_e}{m_{\tilde L_L}^2}\tan\beta \frac{\text{Im}(M_2 \mu)}{|M_2|^2-|\mu|^2}\left[ I_2\left(\frac{M_2^2}{m_{\tilde L_L}^2}\right)-I_2\left(\frac{\mu^2}{m_{\tilde L_L}^2}\right)  \right]\,,
\label{fsusy}
\ee
where  $I_2$ is  defined in \eq{eq:R_2_EDM_full}. These effects are   generated at $\Lambda\sim $ mass of the superpartners.
Taking $\tan\beta \sin ({\rm Arg}[\mu M_2])\sim 1$,
we get  from the new bound \eq{expbound}
\be
m_{\tilde L_L}\gtrsim  25\, (50)\ {\rm TeV}\,,
\ee
 for $m_{\tilde L_L}=M_2=\mu$ ($m_{\tilde L_L}\gg \mu=M_2$).

\begin{figure}[t]
\centering
\includegraphics[width=.4\textwidth]{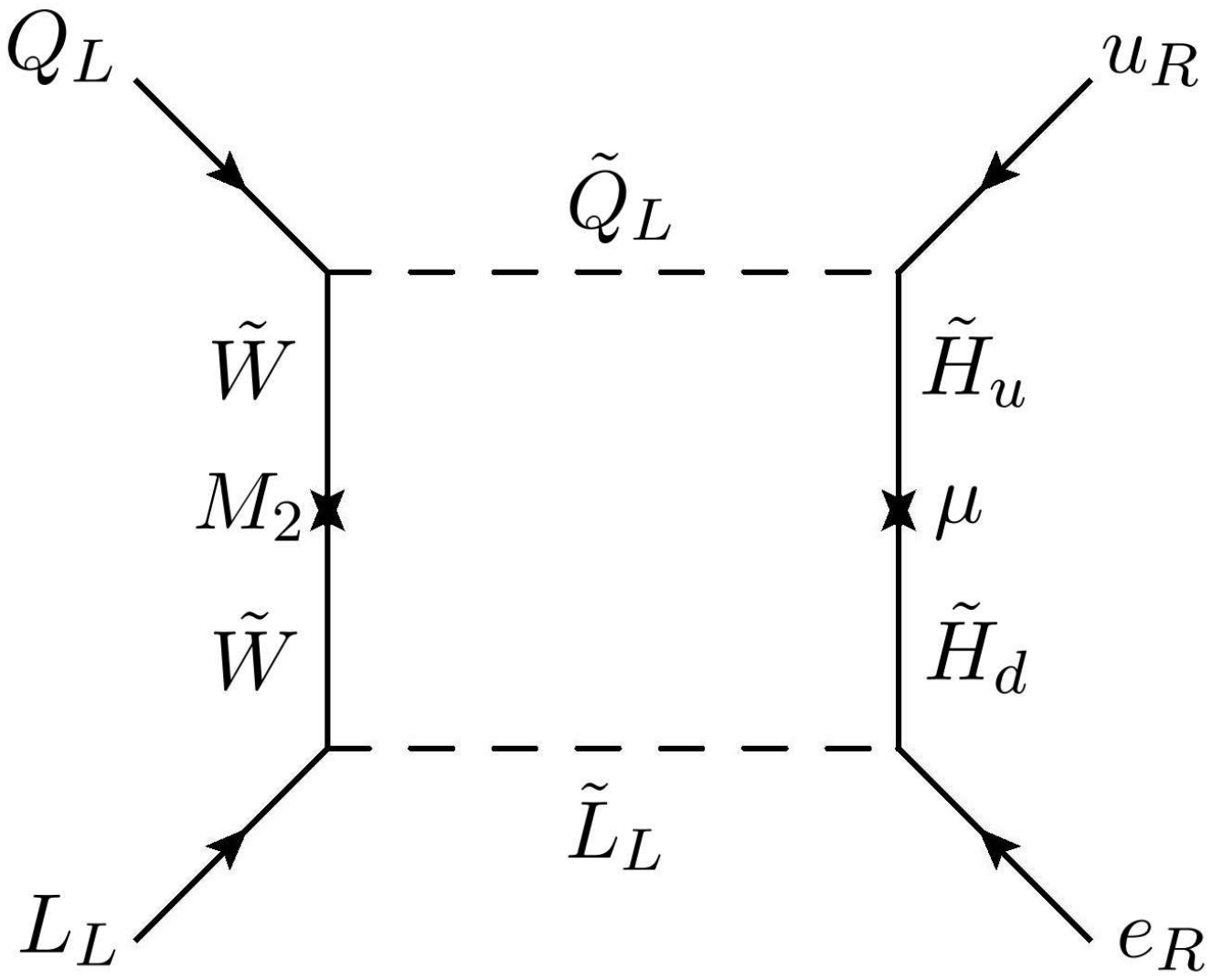}
\caption{\it Feynman diagram of the MSSM contribution to ${\cal O}_{luqe}$.}
\label{feynmanmssm}
\end{figure}

At the two-loop order,  we can get contributions from  other regions of the parameter space of the MSSM.
For example,   Wino-Higgsino loops can induce the Wilson coefficients 
$C_{W\widetilde W},C_{B\widetilde B},C_{W\widetilde B}$ (the Bino contribution is much smaller for $M_1\sim M_2$),
that, contrary to the one-loop \eq{fsusy}, do not involve a $\tilde L_L$.
These are give by
\bea
C_{W\widetilde W}&=&C_{ loop}\,\frac{-8 + 27 \rho - 24 \rho^2 + 5 \rho^3 + 
 6 \rho^2 \ln\rho}{16 (\rho-1)^3}\,,\nonumber\\
 C_{B\widetilde B}&=&t_{\theta_W}^2C_{ loop}\,\frac{\rho (11 - 16 \rho + 5 \rho^2 - 2 ( \rho-4) \ln\rho)}{16 ( \rho-1)^3}\,,\nonumber \\
 C_{W\widetilde B}&=&t_{\theta_W}C_{ loop}\,\frac{\rho (7 - 8 \rho + \rho^2 + 2 ( \rho+2) \ln\rho)}{8 ( \rho-1)^3}\,, 
\label{loopsusy}
 \eea
where $\rho\equiv |M_2/\mu|^2$ and
\be
C_{loop}\equiv \frac{g^4 \sin 2\beta \sin\varphi}{16\pi^2|M_2\mu|}\ ,\ \ \varphi={\rm Arg}[m^2_{12}\mu^* M_2^*]\,.
\label{loopsusy2}
\ee
Using \eq{loopsusy} and \eq{total},
  we can  obtain the contribution to the electron EDM in agreement with Ref.~\cite{Giudice:2005rz}
in the large log-approximation.
From  the ACME bound \eq{expbound},  we get a  limit on the Wino and Higgsino masses  that can be approximately 
written as 
\be
\sqrt{|M_2\mu|}\gtrsim 4\, {\rm TeV}\,,
\ee
where we have taken $\sin 2\beta\sim \sin\varphi\sim 1$.

Another type of two-loop contributions to the electron EDM can arise from one-loop  contributions to $C_{luqe}$.
From  loops involving  selectrons, squarks, Winos and Higgsinos (see Fig.~\ref{feynmanmssm}), we have
\be
{\rm Im}\, C_{luqe}= -y_ey_u\frac{3g^2\, {\rm Im}[\mu M_2]}{16\pi^2\sin 2\beta }F(m^2_i)\,,
\label{cSUSY}
\ee
where
\be
F(m^2_i)=-\sum_i\frac{m^2_i\ln m^2_i}{\Pi_{i\not=j} (m^2_i-m^2_j)},
\ee
 with $i$ running  over the mass of the Higgsino, Wino, $\tilde u_R$ and $\tilde L_L$.
These results are valid for any quark generation $u\to u,c,t$.
For equal superpartner masses, we have $F(m^2_i)=1/(12m_i^4)$, and 
\eq{deluqe} and the ACME bound lead
to 
\be
m_i\gtrsim 7.5\, {\rm TeV}\,,
\ee
for $ \sin(  {\rm Arg}[\mu M_2])/\sin 2\beta\sim 1$.
Notice that, contrary to  \eq{fsusy} and \eq{loopsusy2},  the $\sin 2\beta$ appears in \eq{cSUSY} in the denominator and therefore becomes larger  for small values of $\tan\beta$.

\subsection{Composite Higgs}

As a last example we consider the class of composite Higgs models. For definiteness we focus on minimal scenarios
based on the $\SO(5) \rightarrow \SO(4)$ symmetry breaking pattern, which gives rise to a single Higgs
doublet~\cite{Agashe:2004rs}. Depending on the implementation of the flavor structure different contributions
to the electron EDM can arise. In models based on the anarchic partial compositeness
paradigm naively extended to both quark and lepton sectors~\cite{anarchic}, large contributions arise
at the one-loop level due to the presence of partners of the SM leptons and/or composite vector resonances.
These contributions are generated at the mass scale of the composite states and can be estimated
as~\cite{KerenZur:2012fr,Panico:2015jxa}
\begin{equation}
\frac{d_e}{e} \sim \frac{1}{8\pi^2} \frac{m_e}{f^2}\,,
\end{equation}
where $f$ denotes the Goldstone Higgs decay constant (or equivalently the scale of spontaneous
$\SO(5) \rightarrow \SO(4)$ symmetry breaking).
The new ACME result implies a severe bound on the compositeness scale
\begin{equation}
f \gtrsim 107\;{\rm TeV}\,,
\end{equation}
which pushes these scenarios into highly fine-tuned territory.

The one-loop contributions to the electron EDM can be efficiently suppressed by either introducing flavor
symmetries (in particular a $\U(2)$ family symmetry involving the light fermion generations~\cite{Barbieri:2012uh})
or  generating the light fermion masses  by a bilinear $\bar ff$ mixing with the strong sector~\cite{Panico:2016ull}.
In both cases the leading corrections to the electron EDM arise at the two-loop level
due to the presence of relatively light fermionic partners of the top quark~\cite{Matsedonskyi:2012ym,Panico:2012uw}.

In a large set of minimal models,\footnote{These models are the ones in which only one $\SO(4)$-invariant effective operator exists which gives rise to the Yukawa couplings. This happens, for instance, in the original
holographic MCHM theories~\cite{Agashe:2004rs,Contino:2006qr}, as well as in ``minimally tuned'' scenarios
with a fully-composite right-handed top quark~\cite{Panico:2012uw}.}
only derivative interactions involving the Higgs and the top partners give rise to CP-violating effects.
Using a CCWZ notation (see Ref.~\cite{Panico:2015jxa} for a review
of the CCWZ formalism), a typical representative of such operators is given by
\begin{equation}\label{eq:d_mu_op}
c_t\, d_\mu^i \overline \Psi_1 \gamma^\mu \Psi_4^i + {\rm h.c.}\,,
\end{equation}
where $d_\mu^i$ denotes the CCWZ $d$-symbol, while $\Psi_{1,4}$ are composite fermions in the singlet and fourplet
$\SO(4)$ representation respectively. The $c_t$ coefficient is in general complex, thus containing a CP-violating phase.

The two-loop corrections to the electron EDM arises from Barr--Zee-type diagrams and contain a leading,
log-enhanced contribution. The origin of the latter can be traced back to a two-step evolution.
At the energy scale of the top partners a finite contribution to the ${\cal O}_{W\widetilde W}$, ${\cal O}_{B\widetilde B}$ and
${\cal O}_{W\widetilde B}$ operators is generated, which then according to \eq{r2}
induces a running for the electron EDM~\cite{Panico:2017vlk}.

As an explicit example we report the results  for the $\bf 14+1$ model with a light $\SO(4)$ fourplet
and a fully composite right-handed top.\footnote{For more details on the model see refs.~\cite{Panico:2012uw,DeSimone:2012fs}.} By integrating out the heavy  top-partners, we obtain,  
at leading order in the $v/f$ expansion,
\begin{equation}
\begin{array}{l}
\displaystyle C_{W\widetilde W} = \frac{N_c g^2}{16 \pi^2} (T^3_u)^2 \frac{c_T}{v}\,,\\
\displaystyle \rule{0pt}{2em} C_{B\widetilde B} = \frac{N_c g'^2}{16 \pi^2} Y^2_Q \frac{c_T}{v}\,,\\
\displaystyle \rule{0pt}{2em} C_{W\widetilde B} = \frac{N_c g g'}{16 \pi^2} (-2 T^3_u Y_Q) \frac{c_T}{v}\,,
\end{array}
\label{ch2loop}
\end{equation}
while at the tree-level, that will be relevant later, we get
\be
C_{y_t}=i\sqrt{2} y_t c_T\,,
\label{topcp}
\ee
where we have defined
\begin{equation}
c_T \equiv \frac{\sqrt{2} v y_{L4} y_{Lt}}{m_T^2} {\rm Im}\; c_t\,,
\end{equation}
with $m_T$  being the mass of the  charged-$2/3$ top partner $T$, and $y_{L4},y_{Lt}$ the mixing of the
$Q_L$ doublet with the composite states as defined in
Ref.~\cite{Panico:2017vlk}.
Notice that the contribution to $C_{y_t}$ is purely imaginary, i.e.~CP-violating.

\begin{figure}
\centering
\includegraphics[width=.5\textwidth]{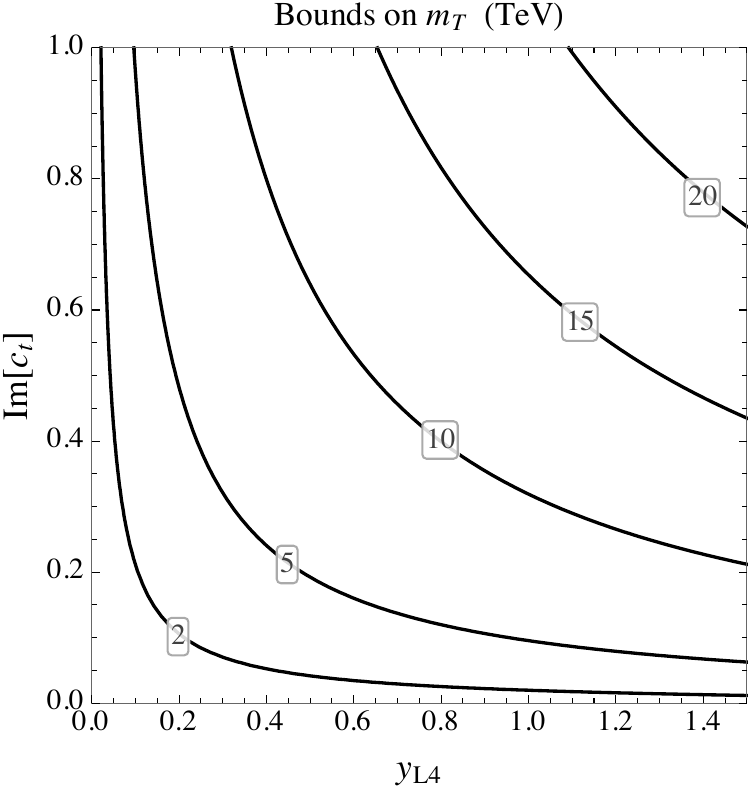}
\caption{\it Bounds on the mass of the  top partner derived from the recent ACME results. The bounds are derived
by setting $f = 800\;$GeV, although the dependence on $f$ is quite mild. The labels on the solid lines show
the top partner mass $m_T$ in TeV.}\label{fig:bounds_ch}
\end{figure}

Using the RGE \eq{total},  we obtain  an electron EDM  given by
\begin{equation}
\frac{d_e}{e} = -\frac{N_c}{64\pi^4} \frac{y_e}{\sqrt{2}} c_T \left[e^2 Q_u^2 + \frac{g^2}{4 c_{\theta_W}^2}Q_u (T^3_u - Q_u s_{\theta_W}^2)(1 - 4 s_{\theta_W}^2)
+ \frac{1}{2} g^2 T^3_u Y_Q\right] \ln \frac{m_T^2}{m_t^2}\,.
\label{chemd}
\end{equation}
The three terms in square brackets come from diagrams containing a virtual photon, a virtual $Z$-boson and a virtual $W$-boson respectively.  Since, as we said,  the $Z$-boson vector coupling to the
electron is quite suppressed,  the  main contribution is  coming from the photon loop, whereas
the $W$-boson term gives a $\sim 40\%$ correction.
In \eq{chemd} the RG running of the EDM starts at  $m_T$ and stops at the top mass.
At that scale, indeed, we have to integrate the top, inducing
an additional finite contribution to $C_{F\widetilde F}$  due to \eq{topcp}.
Surprisingly,\footnote{As noticed in Ref.~\cite{Panico:2017vlk}, the cancellation of the contributions to the ${\cal O}_{F\widetilde F}$ operators at low energy is a direct consequence of the fact that the derivative Higgs operator in Eq.~(\ref{eq:d_mu_op})
induces purely off-diagonal couplings with the composite fermions. For this reason the trace of the coupling matrix
vanishes and the top loop exactly cancels the contributions from the top partners. The cancellation is rather generic
and happens in a large class of models. Indeed, since the $d_\mu$ CCWZ symbol
transform non-trivially under $\SO(4)$ (it is in the representation $\bf 4$), it can only give rise to couplings involving
fermions in two different $\SO(4)$ representations, which are therefore purely off-diagonal.}
 this contribution exactly cancels the one coming from top partners loops, \eq{ch2loop},
so that no net contribution to  ${\cal O}_{F\widetilde F}$ is left below $m_t$.

In Fig.~\ref{fig:bounds_ch} we give the constraints on the mass of the $T$ top partner in the $(y_{L4}, {\rm Im}[c_t])$
plane. To derive these bounds we assumed that the result in \eq{chemd} provides the main correction to
the electron EDM and no additional contributions (or at least no strong cancellations) are present. We can see that for
natural values of the parameters of the theory, $y_{L4} \sim 1$ and ${\rm Im}[c_t] \sim 1$, the bounds from
the electron EDM measurements can exclude top partner masses up to $\sim 15\;$TeV. This bound is significantly stronger
than the current direct LHC exclusions and cannot be matched even in the high-luminosity LHC
runs~\cite{Matsedonskyi:2015dns}.

\section{Conclusions}

We performed a two-loop analysis of the EDM of the electron using the EFT approach.
In particular, we calculated the RGEs of the dimension-6 CP-violating  dipole operators 
at the two-loop order,\footnote{We have only calculated the leading effect to the EDM for each
Wilson coefficient $C_i$ up to the two-loop order. This means that we have not included, for example,
 self-renormalization effects, neither  two-loop effects 
 from Wilson coefficients entering in the renormalization of the EDM at the one-loop level. 
 These effects  are only expected to correct the derived bounds by less than $O(1)$,  and could be easily incorporated if needed (see for example \cite{Czarnecki:2002nt} for the case of the magnetic dipole moment of the muon).  }
as well as the one-loop RGEs of most relevant dimension-8 operators.
We have shown that, due to selection rules,  few operators can mix with  EDM operators, as 
appreciated in  Figure~\ref{fig:2-loop_running} 
where we present a  summary of which and how  dimension-6 operators  enter into the EDM.
We also commented on the RG running of the Wilson coefficients below the electroweak scale, and when CP-violating
electron-nucleon interactions can be competitive with bounds on the electron EDM.

These results are important to provide a proper interpretation  of the new ACME bound   
on the electron EDM in terms of  constraints on BSM particles.
The recent improvement on the bound allows to constrain TeV new-physics
even when it only contributes at the two-loop level. 
 We have shown this with some examples.
In particular, we considered theories with leptoquarks or extra Higgs,
 obtaining bounds ranging  $1-100$ TeV.
We also  considered two of the most motivated BSM scenarios for TeV new-physics, 
supersymmetry and composite Higgs. 
We first reinterpreted  previous calculations in the EFT language.
Then, we used our RGE two-loop results to understand which sectors or which new regions of the parameter space of these BSM are now constrained by the recent ACME result.
 In the MSSM case, for example, we showed how our two-loop results 
can provide new constraints on the small $\tan\beta$ region in the  s-electron and wino sector.
For the composite Higgs, after reinterpreting  calculations 
 on top-partners in the EFT language,
we  showed that  bounds on these particles put them out of the reach of the LHC, 
unless they have CP-conserving couplings.

Therefore, we conclude that,  unless we find a reason of why, contrary to the SM, the interactions in  these BSM do respect CP, the ACME result makes these theories  much less natural.
More importantly, future improvement on the the electron EDM bound  (see for example \cite{Cairncross:2017fip})
could constrain BSM beyond the reach of future colliders.

\medskip
\section*{Acknowledgments}
We thank G.~M.~Pruna and L.~Vecchi for useful discussions.
A.P. has been partly supported by the Catalan ICREA Academia Program, and  grants FPA2014-55613-P, FPA2017-88915-P,  2014-SGR-1450 and Severo Ochoa excellence program  SEV-2016-0588.

\appendix

\section{EDM contributions from Barr--Zee diagrams}\label{app:Barr--Zee}

In this appendix we report the full expressions for the contributions to the electron EDM coming from CP-violating
Higgs Yukawa's originating from $H^3 \bar f f$ operators.

Before giving the formulae for the EDM contributions, it is useful to introduce a few definitions. We define the
$j(r, s)$ function as~\cite{Nakai:2016atk}
\begin{equation}
j(r,s) \equiv \frac{1}{r-s} \left(\frac{r \ln r}{r-1} - \frac{s \ln s}{s-1}\right)\,.
\end{equation}
The Barr--Zee results with a CP-violating quark Yukawa can be rewritten in terms of the following integrals~\cite{Brod:2013cka}
\begin{eqnarray}
F_1(a, 0) &\equiv& \int_0^1 dx \frac{1}{x(1-x)} j\left(0, \frac{a}{x(1-x)}\right)\nonumber\\
&=& \frac{2}{\sqrt{1- 4 a}}
\left[{\rm Li}_2\left(1 - \frac{1-\sqrt{1-4a}}{2 a}\right) - {\rm Li}_2\left(1 - \frac{1+\sqrt{1-4a}}{2 a}\right)\right]\,,
\end{eqnarray}
and
\begin{equation}
F_1(a, b) \equiv \int_0^1 dx \frac{1}{x(1-x)} j\left(b, \frac{a}{x(1-x)}\right) = \frac{1}{1 - b} \left[F_1(a, 0) - F_1(a/b, 0)\right]\,.
\label{eq:appF1}
\end{equation}
Notice that the result in \eq{eq:appF1} follows immediately from the relation $j(r,s) = [j(0,s) - j(0, s/r)]/(1-r) $.
For the diagrams involving a CP-violating electron Yukawa we need the following integrals
\begin{equation}
F_2(a, 0) \equiv \int_0^1 dx \frac{(1-x)^2 + x^2}{x(1-x)} j\left(0, \frac{a}{x(1-x)}\right)
= (1- 2 a) F_1(a,0) + 2 (\ln a + 2)\,,
\end{equation}
and
\begin{eqnarray}
F_2(a, b) &\equiv& \int_0^1 dx \frac{(1-x)^2 + x^2}{x(1-x)} j\left(b, \frac{a}{x(1-x)}\right)\nonumber\\
&=& \frac{1}{1 - b} \left[(1- 2 a) F_1(a,0) - \left(1- 2 \frac{a}{b}\right) F_1\left(\frac{a}{b},0\right) + 2 \ln \frac{a}{b}\right]\,.
\end{eqnarray}
The expansion of the $F_1(a,0)$ integral for large $a$ is given by
\begin{equation}
F_1(a,0) \simeq \frac{1}{a} \left(2 + \ln a\right)\,,
\end{equation}
while for small $a$ one finds
\begin{equation}
F_1(a,0) \simeq \ln^2 a + \frac{\pi^2}{3}\,.
\end{equation}

We can now give the results for the Barr--Zee contributions to the electron EDM. The contribution from a
${\cal O}_{y_f}$ operator for a generic quark or a heavy lepton is given by
\begin{equation}
\frac{d_e}{e} = \frac{N_c}{64 \pi^4} Q_f \frac{m_f m_e}{m_h^2} v \frac{{\rm Im}\, C_{y_q}}{\sqrt{2} \Lambda^2}
\sum_{V=\gamma, Z} g_e^V g_f^V F_1\left(\frac{m_f^2}{m_h^2}, \frac{m_V^2}{m_h^2}\right)\,,
\end{equation}
where $g_{e,f}^V$ denote the vector couplings of the gauge boson $V$, namely
$g_f^\gamma = e Q_f$ for the photon and $g_f^Z = g/(2 c_{\theta_W}) (T^3_f - 2 Q_f s_{\theta_W}^2)$.
The contribution from the ${\cal O}_{y_e}$ operator is instead
\begin{equation}
\frac{d_e}{e} = \frac{N_c}{64 \pi^4} Q_t \frac{m_t^2}{m_h^2} v \frac{{\rm Im}\, C_{y_e}}{\sqrt{2} \Lambda^2}
\sum_{V=\gamma, Z} g_e^V g_t^V F_2\left(\frac{m_t^2}{m_h^2}, \frac{m_V^2}{m_h^2}\right)\,,
\end{equation}
where we only included the contributions from top loops, since the ones from the other fermions are suppressed
by the small Yukawa's.


\end{document}